\documentclass[twocolumn, tighten]{aastex631}

\usepackage[utf8]{inputenc}
\usepackage{amsmath, mathtools, mathrsfs}
\usepackage{hyperref}
\usepackage{verbatim}
\usepackage{xspace}
\usepackage{longtable}
\usepackage{multirow}
\usepackage{afterpage}
\usepackage{booktabs}
\usepackage{bm}
\usepackage{dsfont}
\newcommand\norm[1]{\left\lVert#1\right\rVert}

\defcitealias{M87EHTCI}{Paper~I}
\defcitealias{M87EHTCII}{Paper~II}
\defcitealias{M87EHTCIII}{Paper~III}
\defcitealias{M87EHTCIV}{Paper~IV}
\defcitealias{M87EHTCV}{Paper~V}
\defcitealias{M87EHTCVI}{Paper~VI}

\def \dd{{\rm d}}
\def\muas{\mu{\rm as}} 
\def\dd{\mathrm{d}}
\def\m87{M87$^*$\xspace}
\def\oj287{OJ\,287\xspace}

\def\comrade{\texttt{Comrade}\xspace}
\def\vida{\texttt{VIDA}\xspace}
\newcommand{\ehtim}{\texttt{eht-imaging}\xspace}

\begin{document}

\title{Hierarchical Interferometric Bayesian Imaging} 

\author[0000-0003-3826-5648]{Paul Tiede}
\affiliation{Black Hole Initiative at Harvard University, 20 Garden Street, Cambridge, MA 02138, USA}
\affiliation{Center for Astrophysics $|$ Harvard \& Smithsonian, 60 Garden Street, Cambridge, MA 02138, USA}

\author[0000-0003-2627-0642]{William Moses}
\affiliation{University of Illinois Urbana-Champaign, IL, United States}

\author[0000-0002-9033-165X]{Valentin Churavy}
\affiliation{Institute of Mathematics, Johannes Gutenberg University Mainz, Staudingerweg 9, 55128 Mainz, Germany}
\affiliation{High-Performance Scientific Computing, Center for Advanced Analytics and Predictive Sciences, University of Augsburg, Germany}

\author[0000-0002-4120-3029]{Michael D. Johnson}
\affiliation{Black Hole Initiative at Harvard University, 20 Garden Street, Cambridge, MA 02138, USA}
\affiliation{Center for Astrophysics $|$ Harvard \& Smithsonian, 60 Garden Street, Cambridge, MA 02138, USA}

\author[0000-0002-5278-9221]{Dominic Pesce}
\affiliation{Center for Astrophysics $|$ Harvard \& Smithsonian, 60 Garden Street, Cambridge, MA 02138, USA}
\affiliation{Black Hole Initiative at Harvard University, 20 Garden Street, Cambridge, MA 02138, USA}

\author[0000-0002-9030-642X]{Lindy Blackburn}
\affiliation{Black Hole Initiative at Harvard University, 20 Garden Street, Cambridge, MA 02138, USA}
\affiliation{Center for Astrophysics $|$ Harvard \& Smithsonian, 60 Garden Street, Cambridge, MA 02138, USA}

\author[0000-0002-6429-3872]{Peter Galison}
\affiliation{Black Hole Initiative at Harvard University, 20 Garden Street, Cambridge, MA 02138, USA}
\affiliation{Department of Physics, Harvard University, Cambridge, MA 02138, USA}
\affiliation{Department of History of Science, Harvard University, Cambridge, MA 02138, USA}

\begin{abstract}
Very long baseline interferometry (VLBI) achieves the highest angular resolution in astronomy. VLBI measures corrupted Fourier components, known as visibilities. Reconstructing on-sky images from these visibilities is a challenging inverse problem, particularly for sparse arrays such as the Event Horizon Telescope (EHT) and the Very Long Baseline Array (VLBA), where incomplete sampling and severe calibration errors introduce significant uncertainty in the image. To help guide convergence and control the uncertainty in image reconstructions, regularization on the space of images is utilized, such as enforcing smoothness or similarity to a fiducial image. Coupled with this regularization is the introduction of a new set of parameters that modulate its strength. We present a hierarchical Bayesian imaging approach (Hierarchical Interferometric Bayesian Imaging, HIBI) that enables the quantification of uncertainty for all parameters. Incorporating instrumental effects within HIBI is straightforward, allowing for simultaneous imaging and calibration of data. To showcase HIBI's effectiveness and flexibility, we build a simple imaging model based on Markov random fields and demonstrate how different physical components can be included, e.g., black hole shadow size, and their uncertainties can be inferred. For example, while the original EHT publications were unable to constrain the ring width of \m87, HIBI measures a width of $9.3\pm 1.3\,\muas$. We apply HIBI to image and calibrate EHT synthetic data, real EHT observations of \m87, and multifrequency observations of \oj287. Across these tests, HIBI accurately recovers a wide variety of image structures and quantifies their uncertainties. HIBI is publicly available in the \comrade VLBI software repository.


\end{abstract}

\keywords{black hole physics --- Galaxy: M87 --- methods: data analysis --- methods: numerical --- submillimeter: imaging}

\section{Introduction} \label{sec:intro}

Very long baseline interferometry (VLBI) produces the highest resolution images among all astronomical techniques, now achieving angular resolutions of $\sim 10\,\muas$ \citep[e.g.,][]{radioastron, M87EHTCI}. However, VLBI does not directly image the on-sky source. A perfect VLBI interferometer samples interferometric visibilities, which are related to the Fourier transform of the corresponding image \citep{TMS}. For instance, visibilities $\tilde{I}(u,v)$ corresponding to the total intensity (Stokes~$I$) image $I(x,y)$ are given by,
\begin{equation}
    \tilde{I}(u,v) = \int I(x,y)e^{2\pi i (ux + vy)}\dd x \dd y.
\end{equation}
Given a pair of sites $a, b$, this implies that the baseline $(a,b)$ measures a single Fourier component of the image, whose position in the Fourier space, $(u,v)$, is given by the projected distance in units of wavelength between the sites. We will denote this Fourier component by $\tilde{I}_{ab} = \tilde{I}(u_{ab}, v_{ab})$. The resolution of the interferometer is, therefore, roughly set by the maximum distance between two sites, 
\begin{equation}\label{eq:resolution}
    \rho_{\rm Dif} \approx 1/u_{\max}.
\end{equation}
VLBI imaging aims to invert the measurement process, removing the impact of instrumental corruption and incomplete Fourier sampling to reconstruct the true on-sky image. As a result, VLBI imaging attempts to rebuild the infinite degree of freedom on-sky intensity map from a finite set of observations.

The primary difficulty in VLBI imaging is that the problem is nonidentifiable. To define identifiability, consider that we have a family of distributions that depends on a set of parameters $\theta$, which we denote $p_\theta$. An example of this is the likelihood distribution $\mathcal{L}(\bm{D} | \bm{\theta})$, where $\bm{D}$ is the hypothetical data we observe. The family of distributions $\theta \to p_\theta$ is identifiable if, for each set of parameters, the resulting likelihood distribution is unique. That is, the mapping from parameters to the likelihood distribution is one-to-one. If this mapping is not unique, the problem is stated to be nonidentifiable. For VLBI imaging, this degeneracy is typically of a form that there exists a smooth transformation of the parameters that leaves the likelihood invariant. As a result, this kind of nonidentifiability implies that repeating an observation infinitely many times will not result in a single image. Instead, the resulting space of images will converge to a lower-dimensional surface in the space of all parameters. This kind of identifiability issue is common in VLBI imaging.

Consider an image $I(x,y)$, together with a second function $\delta(x,y)$ that is zero-mean and that has a characteristic scale of variation far below the resolution of the VLBI interferometer (\autoref{eq:resolution}). In this case, $I(x,y) + \delta(x,y)$ will produce an identical likelihood even though we have changed the image potentially everywhere. For posterior inference this means that naive estimation of the image will result in a non-normalizable distribution. That is, there are infinitely many variations $\delta$ that will produce identical fits to the data. To limit the space of images, we must make additional assumptions about the source structure. For example, we could consider a simple parametric model, such as simple geometric shapes \citep[][hereafter \citetalias{M87EHTCVI}]{M87EHTCVI}, if we have a priori knowledge of what we are analyzing. However, for VLBI, this is often not warranted. Simplified geometric models typically underfit the data, since VLBI images describe complicated environments, e.g., turbulent jets, that are difficult to know a priori. The other, more common way to solve this problem in VLBI imaging is to focus on non-parametric models and then regularize the space of images to make it finite and at least weakly identifiable. In VLBI imaging, this is typically achieved through regularization of the space of images. If we focus on forward modeling of the source, this implies the use of Bayesian priors or similar regularizers within the \textit{regularized maximum likelihood} (RML) methods that have recently become popular in VLBI imaging \citep{NarayanMEM, Chael_2018, TSV, mpol}. The goal of these regularizers is to modify the resulting probability distribution such that the map is now identifiable, although this is often difficult to achieve in practice.

However, pure imaging is not the only source of non-identifiability; when instrumental effects are considered, additional degeneracies occur. For a realistic interferometer the observed visibilities are corrupted by both baseline-dependent thermal noise effects ($\epsilon_{ab}$) and station-based complex gains ($g_{a}$), giving
\begin{equation}
    \hat{V}_{ab} = g_{a}\tilde{I}_{ab}g_{b}^* + \epsilon_{ab},
\end{equation}
where $\epsilon_{ab}$ is drawn from a complex Gaussian with real and imaginary standard deviation, $\sigma_{ab}$.
In this paper, we will parameterize $\bm{g}$ with a set of parameters that we will collectively refer to with $\bm{s}$. Namely, we will typically decompose $g$ into
\begin{equation}\label{eq:gain}
    g_a = \exp(\gamma_a + i \varphi_a),
\end{equation}
where $\gamma_a$ and $\varphi_a$ are real numbers and are typically denoted by log gain amplitudes and gain phases, respectively.

The VLBI likelihood for a single frequency and time is given by
\begin{equation}
\begin{aligned}
    \mathcal{L}(\hat{V}_{ab} &| \bm{r}, \bm{s}_a, \bm{s}_b) =\\ &(2\pi \sigma_{ab}^{2})^{-1}\exp\left(-\frac{|\hat{V}_{ab} - g_{a}(\bm{s}_a)\tilde{I}_{\bm{r}, ab}g_{b}^*(\bm{s}_b)|^2}{2\sigma_{ab}^2}\right).
\end{aligned}
\end{equation}
Assuming each time and frequency measurement is independent, the total likelihood is given by
\begin{equation}\label{eq:vlbi_lklhd}
    \mathcal{L}(\hat{\bm{V}} | \bm{r}, \bm{s}) = \prod_{t, a, b} \mathcal{L}(\hat{V}_{t,ab} | \bm{r}, \bm{s}_{t,a}, \bm{s}_{t,b}),
\end{equation}
where $a$, $b$ denote the baselines and $t$ the observation times.

For high frequency VLBI, these gains are typically only known to within $10-20\%$ and can have order-unity uncertainties for some sites due to technical issues or poor weather. Therefore, these gains are estimated during imaging in a procedure typically denoted iterative self-calibration \citep{selfcal_wilkinson_1977, selfcal_cornwell_1981}. The need to jointly estimate $I(x,y)$ and $g_a$ induces additional nonidentifiability issues. For example, a shift in the image centroid $I(\bm{x}) \to I(\bm{x}+\bm{c})$ can be absorbed in each telescope gain phase, meaning that the VLBI likelihood cannot constrain the absolute image position. Similarly, the change in the total flux of the image degenerates with the scaling of the gain amplitude of each telescope. Moreover, while these degeneracies are in principle easy to eliminate, more complicated degeneracies can occur, depending on the telescope performance and coverage of the array, making them impossible to eliminate a priori. 

Given that completely eliminating these degeneracies is extremely difficult, generic VLBI imaging must be able to explore this space of degeneracies and quantify the resulting uncertainty. In that vein, we seek a method that allows us to quantify and even parameterize the space of potential images, and together solve for both the image and instrumental response. To regularize the space of images, we first notice that regularizing the space of images, it is often natural to decompose the set of unknown parameters into different components. 

One component is parameters directly related to the image intensities, which we will denote generically as $\bm{r}$. These could be, for instance, the specific intensities for a rasterized image. The second set of parameters, which we denote by $\bm{\phi}$ encode global properties of the image or parameters of the regularization of $\bm{r}$. Such parameters include the image correlation length, image variation, image size, image shape, etc. Following the language of statistical inference, we call $\bm{r}$ the latent variables and $\bm{\phi}$ the parameters or hyperparameters of the model. The space of potential images we wish to explore is then given by the joint prior
\begin{equation}\label{eq:hier_prior}
    p(\bm{r}, \bm{\phi}) = p(\bm{r} | \bm{\phi})p(\bm{\phi}),
\end{equation}
where the right hand side utilizes Bayes' rule to explicitly denote the conditional dependence of the latent variables on the global parameters of the model. This kind of prior is typically called a hierarchical Bayesian prior, and reflects the natural hierarchy of knowledge that is typical in scientific inference. 

Within the Bayesian framework, adding the instrumental response is straightforward: include the gains as parameters with priors in the model. Putting the different model components together, the total imaging posterior becomes, 
\begin{equation}
    p(\bm{r}, \bm{\phi}, \bm{s} | \bm{V}) = \frac{\mathcal{L}(\bm{V} | \bm{r}, \bm{\phi}, \bm{s})p(\bm{r} | \bm{\phi})p(\bm{\phi})p(\bm{s})}{p(\bm{V})}.
\end{equation}
We call this formulation of VLBI imaging \textit{hierarchical interferometric Bayesian imaging} (HIBI). The benefit of this approach is that depending on the problem at hand we can assess a large variety of different problems that arise in VLBI inference. For instance, for standard imaging problems our main goal is to compute averages or expectations of the image structure. This is typically expressed in terms of the marginal posterior $p(\bm{r} | \bm{V})$, which averages over both the different parameters and instrumental effects. That is, we are averaging over the impact of different imaging priors and their related instrumental effects. In this setting, $\bm{\phi}$ and $\bm{s}$ serve as nuisance parameters whose uncertainty we do want to consider but whose values are not of direct interest to the scientific output.

Conversely, suppose that we are modeling the image as some average feature, i.e. the average state of an accretion flow, plus a turbulent process. Within this setting $\bm{r}$ would represent the specific realization of the turbulent field, and $\bm{\phi}$ the physical parameters of interest, e.g., black hole mass, ring size, black hole spin. The distribution of interest in this case would be $p(\bm{\phi} | \bm{V})$ and so $\bm{r}$ would be the nuisance parameters that we average over. HIBI provides direct uncertainty quantification in both instances and in this paper we will explore both perspectives.

This paper is organized as follows. In \autoref{sec:image_model}, we present a specific implementation of HIBI in terms of Markov random fields and then relate it to other imaging techniques. In \autoref{sec:synthetic_data}, we test HIBI on a variety of image structures based on the 2017 EHT observations of \m87 \citep[][hereafter \citetalias{M87EHTCIV}]{M87EHTCIV} and apply HIBI to the 2017 EHT \m87 data reproducing the results from \citetalias{M87EHTCIV}. In \autoref{sec:ring}, we demonstrate how additional physical information, e.g., the expected appearance of an optically thin accretion flow, can be incorporated into the HIBI framework and reduce the uncertainty of image parameter estimates. In \autoref{sec:AGN} we apply HIBI to the AGN \oj287 across three frequencies independently, demonstrating HIBI's super-resolution capabilities. Finally, in \autoref{sec:summary} we summarize our results and discuss future work.

\section{Hierarchical Bayesian Imaging}\label{sec:image_model}
The locations of the pixels are given by
\begin{equation}
\begin{aligned}
    x_i &= \mathrm{FOV}_x\left[-\frac{1}{2} + \frac{1}{N_x}(i-1/2)\right]\\
    y_j &= \mathrm{FOV}_y\left[-\frac{1}{2} + \frac{1}{N_y}(j-1/2)\right],
\end{aligned}
\end{equation}
where $\mathrm{FOV}_{x,y}$ is the field of view of the image and $i,j$ go from $1$ to $N_x$ and $N_y$ respectively. Note that  $\Delta(x,y) = {\rm FOV}_{x,y} / N_{x,y}$ are the pixels sizes. Associated with each pixel is a specific intensity, $F_{ij}$, whose collection we denote by $\bm{F}$. To create a continuous representation of the image, we use the formula
\begin{equation}\label{eq:image_model}
    I_{\bm{F}}(x,y) = \sum_{i=1}^{N_x}\sum_{j=1}^{N_y} F_{ij} \kappa\left(\frac{x - x_i}{\Delta x}, \frac{y - y_j}{\Delta y}\right),
\end{equation}
where $\kappa(x,y)$ is a continuous function often called the \textit{kernel} or \textit{pulse} function. 

The Fourier transform of \autoref{eq:image_model} is given by
\begin{equation}\label{eq:img_vis}
    \tilde{I}_{\bm{F}}(u, v) = \tilde{\kappa}(u\Delta x, v\Delta y) \Delta x \Delta y\sum_{ij}F_{ij} e^{2\pi i (ux_{i} + vy_{j})},
\end{equation}
where $\tilde{\kappa}$ is the Fourier transform of $\kappa$. To efficiently compute the visibilities for a set of u-v locations, we use a non-uniform fast Fourier transform implemented within the \texttt{NFFT.jl} package \citep{NFFTjl}. For our pulse function, we use the third-order B-spline $b_3(x)$, which is given by the square wave pulse,
\begin{equation}
    b_0(x) = \begin{cases}
                1  & -1/2 < x <1/2 \\
                0  & \rm otherwise
            \end{cases},
\end{equation}
convolved with itself three times.

\subsection{Prior Properties of the Stochastic Process}
Different image priors may give qualitatively different results. Therefore, we aim to construct a prior that encodes the minimal properties we expect our images to obey. For a general image $I$, two generic properties should be considered:
\begin{itemize}
    \item \textit{Intensities are positive}: $I(x,y) \geq 0$ for all $(x,y)$. For our raster, $F_{ij} \geq 0$ for all $i,j$.
    \item \textit{Spatial Correlation}: Neighboring pixels should have similar fluxes. 
\end{itemize}
Positivity is enforced because each $F_{ij}$ measures an energy flux density. The inclusion of spatial correlation in the model is intended to enforce the idea that astrophysical images are often correlated at some spatial scale. Previous Bayesian imaging methods \citep{BroderickHybrid, dmc} have assumed uncorrelated image priors\footnote{The Dirichlet parameterization used in \citet{dmc} could include local correlation in the concentration parameter matrix.}. As a result, the number of pixels becomes an important hyperparameter that can influence the structure of the resulting image posterior. If too few pixels are used, the model underfits the data, often leading to overly tight posteriors. At the same time, too many pixels cause overfitting of the data, resulting in an overly broad posterior and weak inference. \citet{BroderickHybrid} used the Bayesian information criterion to find the optimal number of pixels. However, estimating the Bayesian evidence for each set of data is computationally expensive. Furthermore, the Bayesian evidence is also sensitive to the choice of priors and may not always choose the most predictive model \citep{Gelman_2021}. 

Using a correlated image prior and fitting for the correlation length, we effectively average over different numbers of effective pixels, finding the optimal configuration in a data-driven manner. As a result, the number of pixels is no longer a critical parameter to consider. Instead, rasterization is a discretization of some underlying continuous model, and as the number of pixels increases, the effects of rasterization decrease. Finally, by specifically introducing a correlation parameter, we can impose a prior that utilizes our prior information that the images should not be dominated by features with a much smaller length scale than what the telescope is sensitive to. For VLBI imaging, it is standard practice to suppress structure on scales below $\rho_{\rm Dif}$.

To parameterize the raster emission, we formulate $F_{ij}$ as
\begin{equation}\label{eq:rastmodel}
    F_{ij} = F_0\frac{\mu(x_i, y_j)\eta_{ij}}{\sum_{ij}\mu(x_i, y_j)\eta_{ij}}.
\end{equation}
Here, $\mu(x,y)$ models the a priori structure of the image assumed before imaging the data, which are modulated by the stochastic fluctuations $\eta_{ij}$. This model implies that the imaging process is a multiplicative stochastic process, where $\mu$ models the geometric mean structure of the process and $\eta$ encodes the correlation structure of the image. Finally, we normalize the process to have total flux density $F_0$ so that the degeneracy between total flux $F_0$ and a total scaling of gain amplitudes is fixed.

The other main degeneracy is the lack of an image centroid constraint. The critical component of this degeneracy is that it strongly correlates the image structure with the gain phases across all sites. To remove this correlation from the gain phases, we first modify \autoref{eq:img_vis} by
\begin{equation}
    \tilde{I}_C = e^{-2\pi i (u x_C + v y_C)}\tilde{I}_{F},
\end{equation}
where $(x_C, y_C)$ is the centroid of $I_{\bm{F}}$. This rescaling renders the visibility phases (and, hence, the gain phases) independent of the image centroid. Note that this does not uniquely identify the center of the image with respect to the raster. However, in practice we find that computing this projection dramatically improves sampling efficiency when estimating the image posterior.

To ensure that $\eta_{ij} > 0$, we first augment these variables using a transformation from Euclidean space to strictly positive values. For low dynamic range image reconstructions, such as the 2017 EHT array, we seek a function that is approximately linear for values $r \gg 0$ and then decays to zero $r \ll 0$. A simple function that satisfies these constraints is the {\rm softplus} function:
\begin{equation}\label{eq:softplus}
    \eta_{ij, {\rm sp}} = {\rm softplus}(r_{ij}) = \log(1 + e^{r_{ij}}).
\end{equation}
For $r_{ij} \gg 0$, it is approximately linear, while for $r_{ij} \ll 0$, it is similar to a decaying exponential. This ensures that $\eta_{ij} > 0$ smoothly transitions to zero as $r_{ij} \to -\infty$, while remaining roughly linear for larger values, i.e., where the image is bright. 

For very high dynamic range images, such as VLBA jet imaging, which we consider in \autoref{sec:AGN}, placing the image fluctuations on a linear scale is often a poor description of the data, e.g., the jet is often orders of magnitude dimmer than the core. Therefore, for these higher dynamic range images, the exponential function is preferable to map between fluxes $F_{ij}$ and $r_{ij}$:
\begin{equation}\label{eq:delta_exp}
    \eta_{ij, \rm sm} = e^{r_{ij}}
\end{equation}
Given the decomposition \autoref{eq:rastmodel}, and our specification of $\eta_{ij}$ in terms of $r_{ij}$, the last component of the model is the prior structure for $\bm{r}$, which we now specify.

\subsection{Correlated Image Prior}\label{sec:gmrf}
Gaussian random fields (GRFs) provide a flexible framework for parameterizing the space of functions, thereby forming a prior on this space. The Gaussian nature of these fields implies that the distribution of functions is completely characterized by the mean function $m(\bm{x})$ and a two-point autocorrelation function $k(\bm{x}, \bm{x}')$. 
Therefore, we assume that stochastic fluctuations of our image, $r_{ij}$, are the discretization of some continuous GRF. We assume that the process is mean zero: $m(\bm{x}) \equiv 0$. Therefore, to specify our GRF, only $k(x_i, x'_j)$ must be specified.

Given a discretization, $\bm{x}_{ij}$, the autocorrelation function becomes a matrix $\Sigma_{ij}$ 
\begin{equation}
\Sigma_{ij, (ij)'} = k(\bm{x}_{ij}, \bm{x}_{(ij)'}),
\end{equation}
and the transformed image fluctuations are a zero-mean multivariate Gaussian 
\begin{equation}
    \bm{r} \sim \mathcal{N}(0, \bm{\Sigma}). 
\end{equation}    
However, choosing an autocorrelation function $k$ is not straightforward. To ensure that $k$ is a valid GRF, we need to ensure that $\Sigma$ will always be a positive definite covariance matrix. One way to simplify the construction of $k$ is to assume that the GRF is stationary. That is, the statistical properties at $\bm{x}$ and at $\bm{x} + \bm{c}$ are identical. This implies that $k(\bm{x}, \bm{x}') = k(\norm{\bm{x} - \bm{x}'})$ and the Wiener-Khinchin theorem then states that $k(r)$ is the Fourier transform of a positive semidefinite function or measure often called the power spectrum or spectral density. Stationary GRFs\footnote{RESOLVE uses a GRF prior on both the image and the power spectrum.} form the basis of the RESOLVE algorithm \citep{resolve}, and have been successfully applied to previous EHT observations \citep{resolve_m87}.

For stationary GRFs on a regular Cartesian grid with periodic boundary conditions, the computational complexity to evaluate the GRF is $\mathcal{O}(K\log K)$. However, a general GRF typically scales $\mathcal{O}(K^3)$ for irregular grids, making it infeasible for large rasters. Another complication in estimating the GRFs is that estimating the correct shape of the power spectrum for small data volume, such as for the 2017 EHT arrays, can be quite difficult \citep{GPPriors}.

Rather than solving for a general power spectrum, in this paper, we restrict our correlated image priors to Gaussian Markov random fields (GMRFs). The basis of GMRFs is that the inverse covariance matrix, or precision matrix $Q$, encodes the condition dependencies between variables. That is, suppose that $r_{ij}, r_{jl}$ are conditionally independent given the rest of the variables $r_{-(ij, jl)}$\footnote{The negative sign means all pixel values except $-ij$} i.e. that 
\begin{equation}
    p(r_{ij}, r_{kl} | r_{-(ij,kl)}) = p(r_{ij} | r_{-(ij,kl)}) p(r_{kl} | r_{-(ij,kl)}).
\end{equation}
Then the resultant precision matrix $Q_{ij, kl} = 0$. For a Markov random field where the pixels will depend only on some small set of neighbors, this implies that the precision matrix will be sparse \citep[see, e.g.,][]{GMRF}.

In general, GMRFs are given by the formula,
\begin{equation}\label{eq:GMRF}
\begin{aligned}
                p(\bm{r} | \bm{\mu}, \bm{Q}) &= \sqrt{\frac{\det \bm{Q}}{2\pi^K}}
            \exp\left(-\frac{1}{2}\mathbf{r}^T \bm{Q}\, \bm{r}\right),
\end{aligned}
\end{equation}
where $Q$ is a sparse precision matrix. GMRF priors have been extensively used in geostatistical and epidemiological \citep[see e.g.,][for a review]{spdereview} studies, form the basis of the popular \texttt{R-INLA} spatiotemporal statistical modeling package \citep{martinsRINLA}, and have already been used in black hole inference by EHT \citep{inoisy, Levis_2021_ICCV}. This paper will consider simple stochastic spatial fluctuations encoded by the precision matrix $\bm{Q}$, which we now describe.

\begin{figure*}[!t]
    \centering
    \includegraphics[width=\textwidth]{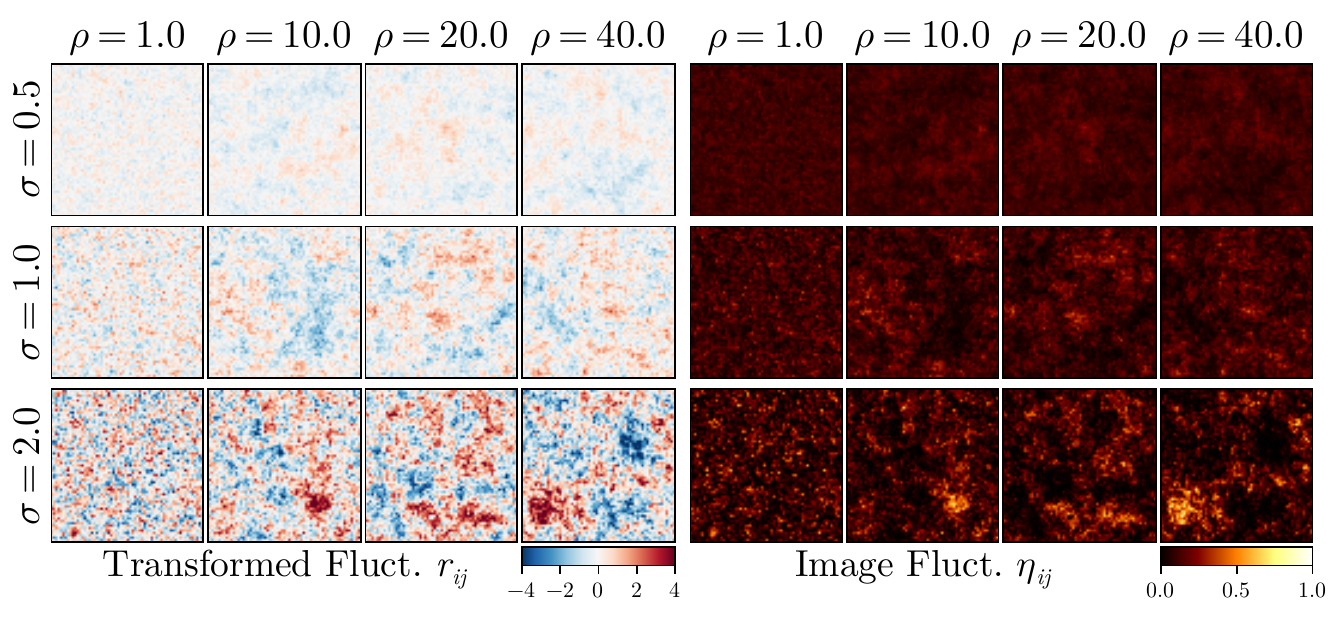}
    \caption{Draws from the GMRF prior in the $\bm{r}$ space (left) and in the linear image space $\bm{\eta}$ (right) assuming that they are related by \autoref{eq:softplus} The columns of the figures show different correlation lengths for the log-ratio image, while the rows are different values of the image dispersion.}\label{fig:0orderGMRF_draws}
\end{figure*}

\subsubsection{First Order Markov Random Field}
For simplicity and to connect with RML imaging, this paper will mainly use a \textit{first-order GMRF}, whose precision matrix is denoted $Q_1$.
To start, we divide our precision matrix into two parts:
\begin{equation}\label{eq:first-order-GMRF}
    \bm{Q}_1 = \frac{1}{\sigma^2}\left( \frac{1}{\rho^2}\mathds{1} + \bm{G}\right),
\end{equation}
where $\mathds{1}$ denotes the identity matrix and $\bm{G}$ sets the correlation structure of the random field. For this work, we followed \citet{SPDE2GMRF} and used the first-order intrinsic GMRF matrix:
\begin{equation}\label{eq:0orderGMRF}
    G_{ij, kl} = \begin{cases}
        4  &  ij=kl\\
        -1 & ij \sim kl\\\ 
        0  & {\rm otherwise}
    \end{cases},
\end{equation}
where $ij\sim kl$ means that on the image pixel grid $ij$ is adjacent to $kl$. With this description, \citet{besag1981system} demonstrated that this 2D GMRF \autoref{eq:first-order-GMRF} converges to a continuous process with spectral density,
\begin{equation}
    S(\bm{k})\dd^2\bm{k} \propto \frac{\dd^2\bm{k}}{1 + \rho^2\norm{\bm{k}}^2}.
\end{equation}
This spectral density implies that the first-order GMRF is a stochastic process with a power law with slope $-1$ and break $\sim\rho$. Therefore, we interpret $\sigma^2$ and $\rho$ as hyperparameters that modulate the marginal variance and correlation length of the process.


 Random draws from the GMRF distribution are shown in \autoref{fig:0orderGMRF_draws} in both $\bm{r}$ and the $\bm{\eta}$ representation.

To generate higher-order GMRF \citet{SPDE2GMRF} showed that by convolving this process with itself $n$ times, it approximates the $n^{\rm th}$-order \textit{Matern} Gaussian process. We briefly overview higher-order GMRFs and Matern processes in \autoref{appendix:higher-order}\footnote{See the Julia package \texttt{VLBIImagePriors.jl} for its open source implementation.}. A benefit of the GMRF approach is that the eigenvalues of $\bm{Q}$ are easily computed in $\mathcal{O}(K)$ time and are given by
\begin{equation}\label{eq:det}
\begin{aligned}
    \det \bm{Q}_1 &= \sigma^{-2K}\prod_{n= 1}^{N_x}\prod_{m=1}^{N_y}(\rho^{-2} + \Lambda_{nm})\\
    \Lambda_{nm} &= 4 + 2\cos\left[\pi n/(N_x+1)\right] + 2\cos\left[\pi m/(N_y+1)\right],
\end{aligned}
\end{equation}
which is derived in \autoref{app:eigenvalues}. This makes computing the GMRF prior $\mathcal{O}(K)$, cheaper than the Fourier representation of generic stationary random fields $(\mathcal{O}(K\log K)\,)$, albeit at the cost of flexibility.


\subsection{Parameter Estimation}\label{sec:hyperparam_selection}

Within the Bayesian interpretation of regularization and GMRF priors, there is a natural and self-consistent way to estimate the image and the ``regularization'' hyperparameters through a hierarchical model. The Bayesian image posterior with the GMRF prior \autoref{eq:flat_0orderGMRF} up to constants is given by
\begin{equation}\label{eq:post_gmrf}
\begin{aligned}
    \log p(\bm{r}, \rho, \sigma| \bm{V}) 
        = &\log \mathcal{L}(\bm{V}| \bm{\theta}) - \frac{1}{2}\bm{r}^T\bm{Q}_1\bm{r} + \\ 
          &\frac{1}{2}\log\det\bm{Q}(\rho, \sigma) + \log p(\rho, \sigma)
\end{aligned}
\end{equation}
where $\mathrm{constants}$ denotes terms that do not depend on any of the posterior parameters, and $\log p(\rho, \sigma)$ is the prior on the imaging hyperparameters. Ignoring the hyperparameter prior and constants, we see that \autoref{eq:post_gmrf} is equal to the negative of \autoref{eq:RML} up to the addition of the log determinant of the precision matrix. Inspecting \autoref{eq:det}, we see that $\det \bm{Q}$ acts as a volume correction that prevents the collapse of the regularizer weights. When the prior is very wide ($\rho$ small and $\sigma$ large), $\det \bm{Q}$ is close to zero. In contrast, as the GMRF prior volume shrinks ($\rho$ large and $\sigma$ small), we get $\det \bm{Q}$ to grow geometrically. Therefore, the Bayesian regularization formulation favors images with large $\rho$ and small $\sigma$, namely simpler images closer to our prior image. This simplicity is then balanced against the data likelihood terms, resulting in a fit that naturally increases the prior volume to account for data complexity.

\subsection{Relating HIBI to Other Imaging Methods}\label{sec:relate}

\subsubsection{RML Methods}\label{ssec:rml}
The usage of priors in Bayesian imaging has been qualitatively compared to RML imaging regularizers, which are commonly used in VLBI \citep{NarayanMEM, Bouman_2016,Chael_2016, Chael_2018, TSV}. This section explores the relationship between the GMRF prior \autoref{eq:0orderGMRF} and two commonly used regularizers. 

RML imaging is a maximum likelihood method that reconstructs a single image given the data by minimizing the cost function
\begin{equation}\label{eq:RML}
    J(\bm{F}; \bm{\lambda}, \bm{D}) = \sum_{d \in |D|} \alpha_d \chi^2_d(\bm{F}) + \sum_{r\in|\lambda|} \lambda_r \mathcal{R}_r(\bm{F}),
\end{equation}
where $|\cdot|$ denotes the set of indices for some tuple, $(\alpha_i, \chi^2_i)$ is the weight and data chi-square of data product $i$ and $(\mathcal{R}_j \lambda_j)$ is the regularizer and its weight respectively. Relating HIBI to RML, it is natural to associate HIBI's data likelihood with the chi-square term in \autoref{eq:RML}, its latent variable prior with the regularizer, and its hyperparameters with the regularizer weights.

For the data terms, HIBI and RML both utilize similar data products; however, from a Bayesian perspective, RML's data chi-squares will modify the weights of various terms. Additionally, RML assumes that each data product is independent, which is true if for instance, closures and visibility are simultaneously included in the data fits, which is common in EHT analyses \citep{M87EHTCIV}. Additionally, the weights $\alpha_i$ are often different from the VLBI likelihood \autoref{eq:vlbi_lklhd}; for example, $\alpha$ is usually given by the inverse of the number of measurements, while the VLBI likelihood omits that factor. Note that while RML only cares about relative weights, Bayesian inference, specifically uncertainty estimates, depends on the absolute scale/weights of all data products.  
Two commonly used regularizers are the \textit{total squared variation} (TSV) and $L_2$ regularizers \citep{TSV} which are given by
\begin{equation}\label{eq:tsv_l2}
    \begin{aligned}
        \mathcal{R}_{\rm TSV}(\bm{F}) &= \sum_{ij} (F_{i+1,j} - F_{i,j})^2 + (F_{i,j+1} - F_{i,j})^2 \\
        \mathcal{R}_{\rm L_2}(\bm{F}) &= \sum_{ij} F_{ij}^2.
    \end{aligned}
\end{equation}

Expanding the negative log density of the GMRF prior (\autoref{eq:GMRF}),\footnote{We are ignoring boundary terms in this equation.} we have
\begin{equation}\label{eq:flat_0orderGMRF}
    \begin{aligned}
        -\log p(\bm{r} | \rho, \sigma) =  \\
            & \frac{1}{2\sigma^2}\sum_{ij} (r_{i+1, j} - r_{i, j})^2 + (r_{i, j+1} - r_{i,j})^2\\  
            & + \frac{1}{2(\rho\sigma)^2}\sum_{ij} r_{ij}^2 \\
            &- \frac{1}{2}\log \det \bm{Q}_1 + \mathrm{constants}
    \end{aligned}
\end{equation}
The first term corresponds to the TSV regularization, while the second is the $\ell_2$ regularization. Furthermore, we can relate the regularization hyperparameters to the standard deviation and correlation length
\begin{equation}\label{eq:hyper2bayes}
    \rho = \sqrt{\frac{\lambda_{\rm TSV}}{\lambda_{\ell_2}}} \qquad \sigma =\frac{1}{\sqrt{2\lambda_{\rm TSV}}}.
\end{equation}
At face value, our prior example is TSV and $L_2$ regularization with a non-standard parameterization of the regularization weights. By letting $\rho\to\infty$, we recover the standard ${\rm TSV}$ regularization.

Other classes of regularizers can also be included in this framework. For instance, one could consider a combination of the \textit{total variation} (TV) and $\ell_1$ regularizers for the log prior:
\begin{equation}
\begin{aligned}
        \log p_{\rm TV}(\bm{r}&| \lambda_{\rm TV}, \lambda_{\ell_1}) = \\&-\lambda_{\rm TV}\sum_{ij}\sqrt{(r_{i,j+1} - r_{i,j})^2 + (r_{i+1, j} - r_{ij})^2}\\
                                   &- \lambda_{\ell_1}\sum_{ij} |r_{ij}| + \log N(\lambda_{\rm TV}, \lambda_{\ell_1}),
\end{aligned}
\end{equation}
where $N$ is the normalization constant. Similar to TSV and $\ell_2$, we can relate the regularization hyperparameters to a pseudo-variance and correlation length using \autoref{eq:hyper2bayes}. Unfortunately, a closed-form expression for $N$ is not known unless $\lambda_{\ell_1} = 0$. As we will see in \autoref{sec:hyperparam_selection}, estimating the hyperparameters in a Bayesian framework requires an expression for the normalization $N$; therefore, we do not consider this prior family in this paper.

Before continuing, we note an important consideration: RML imaging tends to regularize the image pixel fluxes directly rather than the transformed quantities. To enforce positivity, the image priors are then effectively truncated multivariate normal distributions. However, this truncation makes the calculation of the normalization term in \autoref{eq:post_gmrf} computationally expensive. Given that this normalization constant is critical to self-consistently estimating the optimal hyperparameters, we do not explore applying the GMRF prior directly to the pixel fluxes. 

Finally, we note that there exists a simple extension of this formalism to multiple regularizers. The standard RML approach of summing multiple regularizers is not easily expressible within HIBI. Unfortunately, the prior volume term in \autoref{eq:post_gmrf} is generally not analytic. However, we can extend the formalism described in \autoref{sec:hyperparam_selection} to multiple priors through model averaging. Let $p_i(\bm{r} | \varphi_i)$, denote each prior considered, with their own set of hyperparameters $\varphi_i$. For each $p_i$, a posterior $p_i(\bm{r}, \varphi_i | \bm{V})$ can be found with the typical HIBI algorithm. These posteriors can be averaged using a model averaging scheme (see, e.g., \citet{modelavg} for different averaging approaches). Using this averaging approach, multiple image reconstruction priors can be combined, weighing each posterior according to the specific averaging scheme. In \autoref{appendix:higher-order}, we describe different Markov random field priors that could be considered in such a scheme.
\begin{figure*}[!t]
    \centering
    \includegraphics[width=\textwidth]{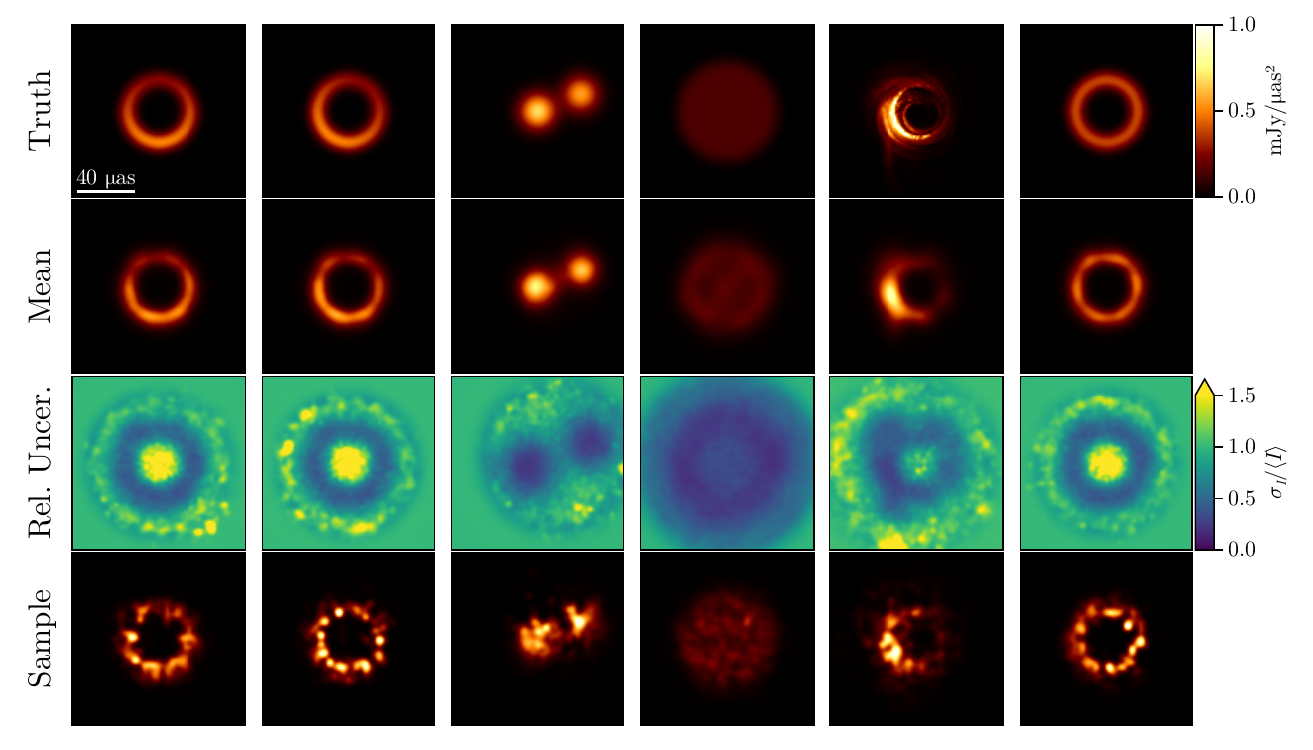}
    \caption{Example reconstructions of various geometric model morphologies using the GMRF prior. The prior is flexible enough to capture various image morphologies and features, including the ring brightness profile.}
    \label{fig:synthetic_data_results}
\end{figure*}

\subsubsection{Comparing RML with HIBI}

Within the HIBI framework, the estimation of the joint image and hyperparameter posterior $\pi(\bm{r}, \rho, \sigma | \bm{V})$ is analogous to the image and hyperparameter survey in RML imaging. A natural question is whether the HIBI formalism could be incorporated into RML imaging by including the missing volume term. The maximum likelihood estimate would then be replaced with the maximum a posteriori (MAP) estimate, but otherwise the rest of the RML algorithm would remain the same. However, as we will show below, the MAP estimate is often a poor approximation of the kind of structure characteristic of the posterior. This is often because the MAP estimate tends to overfit the data and results in various image artifacts. However, another problem exists with the MAP estimates in the HIBI imaging framework: the MAP solution is not invariant to parameterizations. Specifically, the HIBI MAP strongly depends on the choice of parameterization, especially for weakly informative data relative to the model complexity, as is often the case for VLBI imaging. 

To demonstrate how the location of the MAP can be changed arbitrarily, we consider the common model known as Neal's funnel. This distribution has the probability density,
\begin{equation}
    p(\bm{x}, v) = \prod_{i=1}^N\mathcal{N}(x_i | 0, e^{v})\mathcal{N}(v | 0, b^2),
\end{equation}
and corresponds to a Gaussian hierarchical model where the variance of the process has a log-normal prior with variance $b$.\footnote{This distribution is equivalent to the VLBI image posterior with the first-order uncorrelated GMRF prior and no observations.}. The log probability density of this model (ignoring constant terms) is
\begin{equation}
    \log p(\bm{x}, v) = -\frac{Nv}{2} - \frac{e^{-v}}{2}\sum_{i=1}^N x_i^2  - \frac{v^2}{2b^2}. 
\end{equation}
The MAP does not exist within this parameterization, but its location tends to $v \to -\infty$, $x_i=0$. Note that this parameterization is similar to the one typically used in RML imaging. 

Now consider the transformed parameters $\tilde{x},\,\tilde{v}$, where $x_i = e^{b/2}\tilde{x}_i + \tilde{x}_0$, $v = b\tilde{v} + \tilde{v}_0$, where $\tilde{x}_0$ and $\tilde{v}_0$ are the effective origin. This parameterization effectively standardizes the distribution, and the probability density becomes the multiplication of two independent unit normal distributions in $\tilde{v}$ and $\tilde{x}$. In this case, the MAP is finite and is given in our original parameters by $x_i = v = 0$. 
By making a relatively simple parameterization change, we shifted the map by an infinite amount. 

This dependence makes the parameterization or coordinate frame a critical piece of HIBI imaging, even though the expectations from the posterior are invariant to the choice of parameterization. As a result, we do not recommend that RML methods switch to the hierarchical prior approach, given that the choice of parameterization, especially for sparse data, can significantly impact the MAP's location and imaging results. Instead, we refer the reader to other methods for hyperparameter optimization, such as the multiobjective optimization approach in \citet{moead}.

\subsection{\texttt{RESOLVE}}
The \texttt{RESOLVE} algorithm is an example of HIBI with an additional hierarchical level. Namely, \texttt{RESOLVE} typically uses \autoref{eq:rastmodel}, typically with $\mu$ flat and the exponential transfer function \autoref{eq:delta_exp} \citep{VLBIResolve}. For the stochastic field \texttt{RESOLVE} use stationary random fields on a periodic regular grid to utilize the power spectrum decomposition. However, instead of assuming some parameteric form for the power spectrum, they then use a second GRF for the power spectrum itself. The prior decomposition equals
\begin{equation}
    p(\bm{r}, \bm{f}, \bm{\phi}) = p(\bm{r} | \bm{f})p(\bm{f} | \bm{\phi})p(\bm{\phi}),
\end{equation}
where $\bm{f}$ are the parameters of the discretized power spectrum and $\varphi$ are the hyperparameters for the stochastic process governing $\bm{f}$. This model is more expressive than the GMRF prior described in \autoref{sec:gmrf}, which assumes a specific power spectrum slope. 

The tradeoff \texttt{RESOLVE}'s expressibility compared to the GMRF approach in this paper comes during inference. Namely, in this paper, we employ Markov Chain Monte Carlo methods to approximate the posterior, whereas \texttt{RESOLVE} typically relies on more approximate parametric methods, such as metric Gaussian variational inference \citet{mgvi}. A downside of the variational approach is that expectations may be poorly approximated if the variational family does not match the posterior. 

Given the sparsity of the EHT data, this paper uses Markov Chain Monte Carlo (MCMC) methods to approximate posterior expectations. Specifically, given that imaging typically requires 1,000 to 10,000 parameters for EHT data, we employ Hamiltonian Monte Carlo and the NUTS algorithm \citep{hoffman_no-u-turn_2011}. To demonstrate the effectiveness of this approach, we now consider a suite of synthetic data tests.


\section{Imaging With HIBI}\label{sec:synthetic_data}
\begin{figure*}[!ht]
    \centering
    \includegraphics[width=\linewidth]{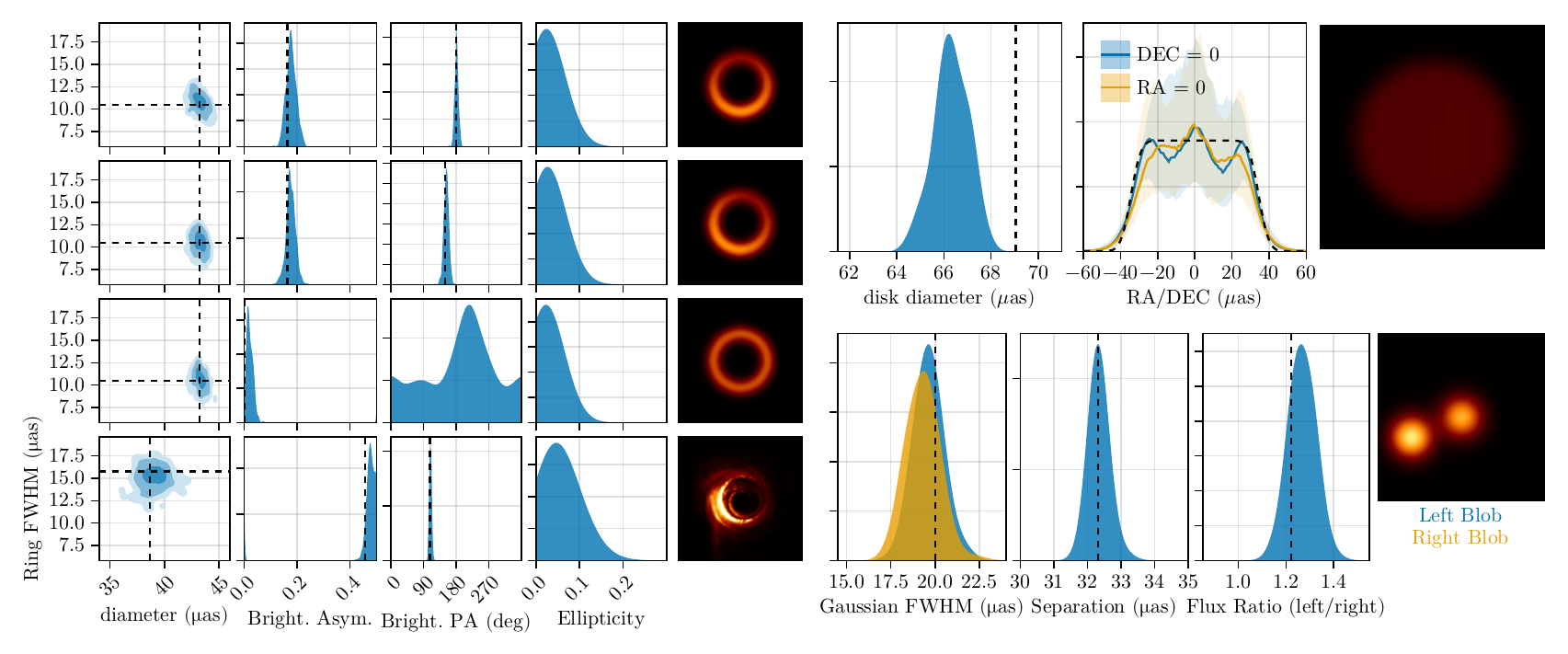}
    \caption{The parameter estimation results from the synthetic data tests in \autoref{fig:synthetic_data_results}. Overall, HIBI can measure all the parameters of the ground truth image except the disk, where the diameter definition gives a slightly too small result. Analyzing the profile of the disk reconstruction, shown between the disk diameter marginal and on-sky truth, we see that at smaller distances from the origin ($< 20~\muas)$, the disk is slightly too dim, while at larger distances ($> 40~\muas$) it is too bright.}
    \label{fig:synth_param_results}
\end{figure*}

\subsection{Synthetic Data Tests}\label{ssec:simdata}
To demonstrate the applicability of our approach to different image source morphologies, we consider a suite of synthetic data that model the observations of the 2017 EHT array following \citet{M87EHTCIV}:
\begin{itemize}
    \item \textit{Disk}: uniform disk with a diameter of $70~\muas$, convolved with a Gaussian with an FWHM of $10~\muas$.
    \item \textit{Double}: Two circular Gaussian components each with a FWHM of $20~\muas$ and a separation of $30~\muas$ RA and $12~\muas$ in DEC. One Gaussian has a flux of 0.33~Jy and the other 0.27~Jy. 
    \item \textit{Ring}: An infinitely thin ring with radius $22~\muas$, flux $0.6~{\rm Jy}$. The ring is convolved with a $10~\muas$ FWHM Gaussian.
    \item \textit{Crescent 150}: An infinitely thin crescent model with radius $22~\muas$, flux $0.6~{\rm Jy}$ and brightness position angle of $150^\circ$, that is blurred with a $10~\muas$ FWHM Gaussian.
    \item \textit{Crescent 180}: An infinitely thin crescent model with radius $22~\muas$, flux $0.6~{\rm Jy}$ and brightness position angle of $180^\circ$, that is blurred with a $10~\muas$ FWHM Gaussian.
    \item \textit{GRMHD}: A general relativistic magnetohydrodynamic (GRMHD) simulation of M87. We use the same simulation as in \citep{M87EHTCIV}.
\end{itemize}
The ground truth images are shown in the first row of \autoref{fig:synthetic_data_results}.
For each image, we then create synthetic data using \ehtim with the EHT coverage and thermal noise properties of the observation of the LO band on 6 April. The synthetic data for each source was also corrupted with random gain phase and gain amplitude errors. In addition, we added a fractional systematic error equal to $2\%$ of the measured visibility amplitudes in quadrature to the thermal noise to all baselines. This additional systematic error models the unresolved errors in the 2017 EHT data, e.g., leakage and coherence loss, that were estimated in \citep{M87EHTCIII}. Note that the same systematic error budget was added to the 2017 EHT data in \autoref{sec:M87}.

\subsection{Model Priors}
For imaging, we used the GMRF prior, with a field of view of $200\,\muas$ with a $64\times64$ raster. A Gaussian profile was used for a priori image structure with the parameterization:
\begin{equation}\label{eq:mug}
    \mu_g(\bm{x}) = \frac{4 \log(2)}{\pi s_g^2}e^{-4\log(2) \bm{x}\cdot\bm{x}/s^2_g}.
\end{equation}
The size or FWHM of the Gaussian profile, $s_g$, is included as a free parameter during inference with the prior,
\begin{equation}
    s_g \sim \mathcal{TN}(50\,\muas, 20\,\muas; a = 20 \muas, b = 100 \muas),
\end{equation}
where $\mathcal{TN}(\mu, \sigma; a, b)$ is a truncated normal distribution with mean, $\mu$, standard deviation $\sigma$, and lower and upper bounds $a$ and $b$, respectively. This prior was chosen to match the estimated size of \m87 based on the analysis from \citetalias{M87EHTCIV}.

For the log-ratio correlation length of the GMRF $\rho$ and dispersion $\sigma$, we used the priors
\begin{align}
    \rho &\sim \mathcal{IG}\left[1, -\log(0.01)\lambda_{\rm EHT}/\Delta x)\right] \\
    \sigma &\sim \mathcal{HN}(0, 0.5^2),
\end{align}
where $\mathcal{IG}(\alpha, \beta)$ denotes inverse-gamma distribution,
\begin{equation}
    p_{\mathcal{IG}}(x |\alpha, \beta) \propto x^{-(\alpha +1)}e^{-\beta/x}
\end{equation}
and $\mathcal{HN}(\mu, \sigma)$ is the half-normal distribution. The parameters for $\rho$ were chosen such that $1\%$ of the prior mass for the correlation length of the MRF was below the telescope beam size.

For our instrument model priors, we used a log-normal prior on the gain amplitudes with a log-mean of zero and a standard deviation of $0.2$ for all baselines except LMT, which assumed a standard deviation of $1$ to model the poor performance of LMT in 2017 observations \citep{M87EHTCIII}. For the gain phases, we used a von Mises distribution,
\begin{equation}
    p_{\rm VM}(\varphi | \mu, \kappa) = 
        \frac{1}{2\pi I_0(\kappa)}e^{\kappa \cos(\varphi-\mu)},
\end{equation}
with mean zero and concentration parameter $\kappa=\pi^{-2}$ for all non-ALMA gain phases. This prior is essentially flat on its support $[0, 2\pi)$ and is temporally uncorrelated. ALMA is used as the reference station of the array, meaning we set its gain phases to $0$ for all scans.

\subsection{Synthetic Data Results}\label{sec:inference}

To reconstruct the image, we used a two-step procedure. The first step is similar to RML, where we found the MAP estimate using the \texttt{Optimization.jl} library and the Adam optimizer from \texttt{Optimisers.jl}. The goal of the first step was to reduce the burn-in time when sampling the posterior, which is the second step. To sample the posterior, we used the \texttt{AdvancedHMC.jl} NUTS sampler with a diagonal mass matrix adaptively tuned over 8,000 MCMC steps. To compute the posterior gradient with respect to the parameters, we used the automatic differentiation software \texttt{Enzyme} and its Julia extension \texttt{Enzyme.jl} \citep{NEURIPS2020_9332c513, 10.1145/3458817.3476165, 10.5555/3571885.3571964}. After tuning, the sampler ran for an additional 7,000 MCMC steps. The runtime of the sampler depended on the dataset, but it generally took 1-2 hours for Stokes I imaging on an AMD Ryzen 7950X CPU using a single core. 

\begin{figure*}[!ht]
    \centering
    \includegraphics[width=\linewidth]{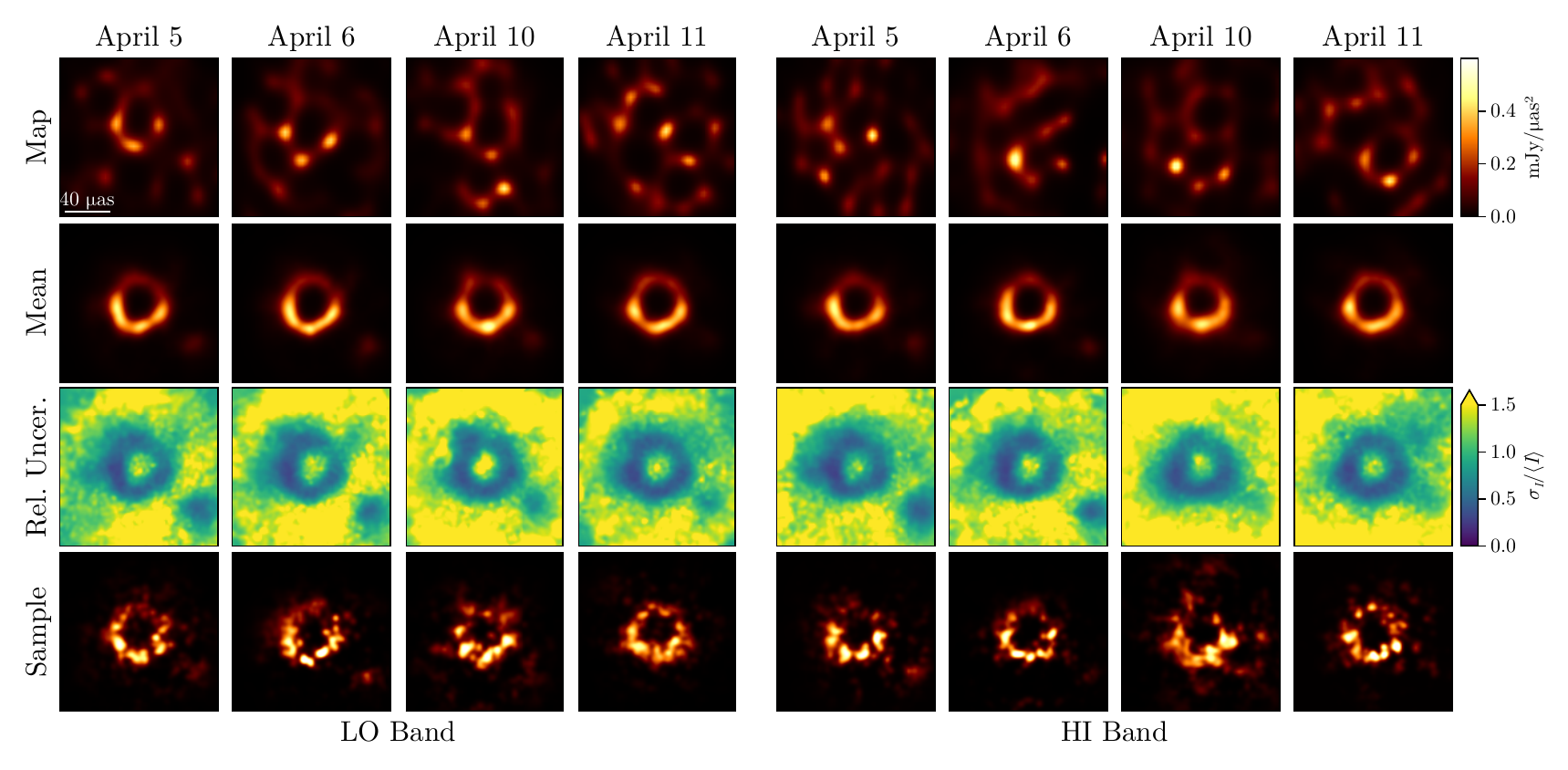}
    \caption{HIBI image reconstructions of \m87 for each observing day and frequency band. The top row shows the MAP image, which is often littered with artifacts, such as ghost rings. The second and third rows display the mean image and its relative uncertainty, respectively. Finally, the bottom row shows a random sample from the posterior, blurred to match the resolution of the mean image using the NXCORR metric.}
    \label{fig:map_v_post}
\end{figure*}

The image reconstructions for each synthetic data test are shown in \autoref{fig:synthetic_data_results}. The second row shows the mean reconstruction, and the third row shows the relative uncertainty of the posterior reconstructions. Qualitatively, we see that HIBI recovers all the different image morphologies considered. 

Note that the disk displays a ring-like inner depression in the mean image. This is only visible in the mean image, and reflects the fact that the central region of the disk is quite uncertain (see \autoref{fig:synth_param_results}. In conjunction with the a priori image being a Gaussian that prefers to concentrate flux near the center, the mean image tends to be brighter near the middle of the disk. A benefit of HIBI's Bayesian approach is that we can assess the significance of this feature and other quantitative features. From the third row of \autoref{fig:synth_param_results}, we see that the relative uncertainty in the central region of the disk is larger than near the edge. Samples from the disk do not consistently show an increased flux near the center, which means that this enhanced brightness near the disk center is not statistically significant.

To quantitatively assess image reconstruction quality, we apply \vida \citep{vida}, a template matching scheme, to extract estimates of key image features. \vida requires a cost function (usually a probability divergence) and a template family to match to the observed image. For the crescent, ring, and GRMHD models, we followed \citet{vidaii} and used an elliptical ring template with a fourth-order cosine expansion for the azimuthal brightness profile. For a definition of the template, see \citet{vidaii} for a complete description. The parameters we report in this paper are:
\begin{itemize}
    \item $d = \sqrt{ab}$ the geometric average of the ellipse semi-major/minor axis of the elliptical ring template
    \item $w$ the FWHM of the Gaussian ring profile
    \item $A$ one half the first order amplitude in the cosine expansion, which we refer to as the \textit{brightness asymmetry}
    \item $\xi$ the first order phase in the cosine expansion, which we refer to as the \textit{ring position angle (PA)}.
\end{itemize}
For the double synthetic data, we used a two-component circular Gaussian blob template, and modeled the two Gaussian blob relative separations and their absolute size. For the disk synthetic data, we used a top-hat disk with a Gaussian taper for the disk synthetic data template \citep[see][for a description]{vida} and only report the diameter of the disk. 

Given a template family, the optimal template parameters are found by minimizing the Bhattacharyya divergence,
\begin{equation}
    {\rm Bh}(t | r) =  -\log  \sum_{n,m}\sqrt{t_{nm}r_{nm}}.
\end{equation} 
To optimize the function, we use the \textit{Evolutionary Centers Algorithm} from the Julia package \textit{Metaheuristics.jl} \citep{metaheuristics2022} within the \texttt{Optimization.jl} package.

The parameter estimation results for all models are summarized in \autoref{fig:synth_param_results}. For each synthetic data test, the true values are within the bulk of the posterior except for the disk diameter. Analyzing the reason for the bias, we plot the true disk profile compared to the recovered profile in \autoref{fig:synth_param_results}. The origin of the bias is that the profile of the disk edge is not accurately reconstructed, especially in the vertical direction, where it is biased to be small near the center and larger near the edge. This is due to the choice of a Gaussian profile $\mu$, which a priori concentrates the emission near the center, as seen in the profile. In \autoref{appendix:ring_synthetic} we demonstrate that a different $\mu$ can remove this bias and recover the true diameter of the disk. Regardless, this bias is minor and generally indicates that HIBI can recover a wide variety of image structures, even when the a priori image structure is not an accurate description. This robustness is critical for arrays such as the EHT, since it provides novel insights into the core region of a variety of AGN \citep{3C279eht, CenAEHT, J1924eht, NRAO530, AGNsurveyEHT} whose structure is highly uncertain a priori.

\subsection{Application to 2017 EHT M87 Data}\label{sec:M87}
\begin{figure*}[!t]
    \centering
    \includegraphics[width=\linewidth]{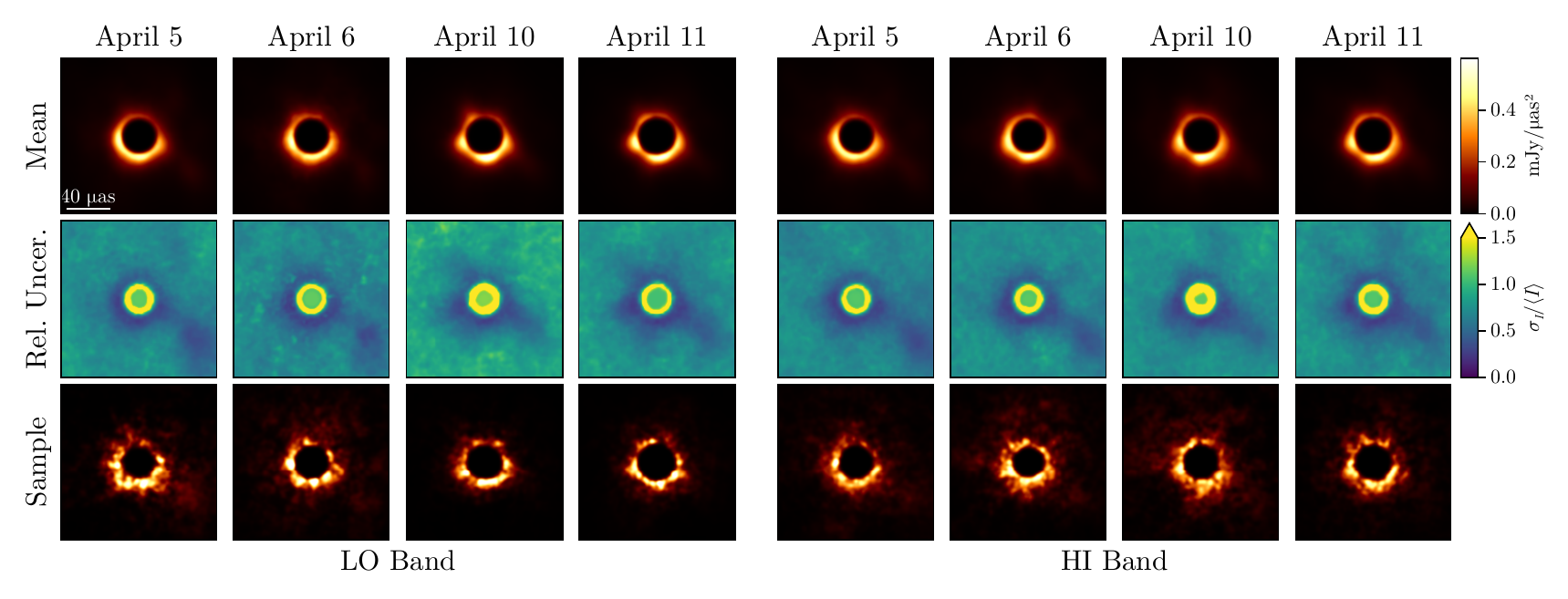}
    \caption{Reconstructions of M87* using the ring profile prior and first order GMRF for multiplicative noise.}
    \label{fig:m87ring}
\end{figure*}

In the previous section, we demonstrated that the GMRF prior and hyperparameter selection algorithms we developed are capable of recovering a wide array of image features from synthetic EHT data. This section applies an identical procedure to the 2017 EHT data\footnote{We used the public EHT data \doi{10.25739/g85n-f134}}. We explore the stability of the image and the properties of the ring.

For the reconstructions of \m87, we followed a similar procedure to \autoref{sec:synthetic_data}. The instrument priors are identical to those for the synthetic data, except on April 11, where a large initial gain in JCMT was found. Therefore, for the data on April 11, we used a log-normal prior with zero mean and a standard deviation of $0.5$ for JCMT.

The imaging results for the 4 days and the two frequency bands are shown in \autoref{fig:map_v_post}. The top row of \autoref{fig:map_v_post} demonstrates the dangers of using a MAP estimate compared to posterior expectations (e.g., the second row of \autoref{fig:map_v_post}). For the imaging results for \m87, the MAP estimate had a reduced chi-square of $\sim 0.4-0.5$, significantly smaller than the reduced chi-square in the posterior bulk ($1.0 \pm 0.1$). This result implies that the MAP estimate overfits the data and produces significant artifacts in the image. Samples from the posterior row in the lower row of \autoref{fig:map_v_post} demonstrate that the MAP is not characteristic of the posterior bulk and should not be used to assess the features of the image when using HIBI.

The second row shows that the mean posterior image is qualitatively similar throughout the four days, as found in \citetalias{M87EHTCIV}. We examined the relative uncertainty map and found that the ring is robustly recovered using HIBI. However, beyond the ring, there is evidence for an extended emission knot in the southwestern region of the image. This extended emission was also found in \citet{resolve_m87, broderick_photon_2022, carilli2022}. 

\section{Parameter Estimation with HIBI}\label{sec:ring}
\begin{figure*}[!ht]
    \centering
    \includegraphics[width=\linewidth]{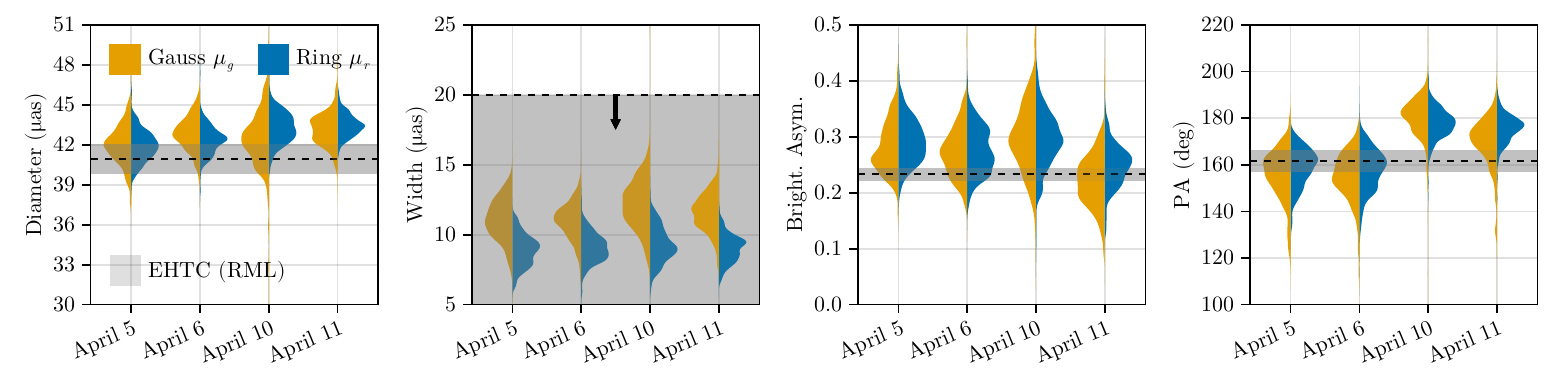}
    \caption{Marginal distribution of the extracted ring parameters from the \m87 LO band data from the Gaussian (orange) and ring (blue) a priori image model. Overall, the reconstructed parameters are similar across the two models, but the ring prior produces tighter measurements of all the ring quantities. The measured values from \citet{M87EHTCIV} averaged over the three separate imaging pipelines are shown by the black lines (mean) and grey band (standard deviation), except for the ring width, where the dotted line and grey region show the upper limit.}
    \label{fig:ring_params}
\end{figure*}

Typically VLBI imaging attempts to be largely agnostic about the on-sky source structure. This approach is ideal when we do not want to make strong structural assumptions about the source, as was the case for the first images of \m87's shadow. However, this approach often ignores our physical understanding of the source and introduces additional sources of uncertainty in our conclusions. Relatedly, being entirely agnostic to the underlying physics often means that physical quantities of interest, e.g., black hole mass and spin, must be extracted, introducing new sources of uncertainty and methodology.

In this section, we take a different perspective and view the image itself as a set of nuisance parameters, while the parameters $\phi$ represent the physical quantities of interest. By using this formulation, we will demonstrate how HIBI can incorporate our knowledge that optically thin accretion flows around supermassive black holes generically produce ring like-morphologies \citep[e.g.,][]{M87EHTCV,SgrAEHTCIV}, and estimate physical parameters of interest directly, e.g., the ring size and its profile. 

We seek to create a simple model that describes the generic parameters of the emission profile around a supermassive black hole. From simulations of accretion flows, these emission profiles often appear to have a profile where the inner profile differs from the outer profile \citep{M87EHTCV, LockhartGrallaII}. A simple model that captures this behavior is the double power-law radial profile 
 \begin{equation}\label{eq:ring_profile}
     \mu_r(\bm{x}| r_0, \alpha_{\rm in}, \alpha_{\rm out}) = N \frac{(r/r_0)^{\alpha_{\rm in}}}{1 + (r/r_0)^{\alpha_{\rm in} + \alpha_{\rm out} + 1}},
 \end{equation}
 Here, $\alpha_{\rm in, out}$ are the inner and outer power-law indices, $r_0$ denotes the rough location of the transition of the power law from the inner to outer scale, and $N$ is a normalization constant to ensure the mean image raster has unit flux. The radius of the peak intensity in \autoref{eq:ring_profile} is given by
\begin{equation}
     r_{\rm peak} = r_0\left(\frac{\alpha_{\rm in}}{\alpha_{\rm out} + 1}\right)^{1/(\alpha_{\rm in} + \alpha_{\rm out} + 1)}.
\end{equation}

To incorporate this model into the HIBI framework, we replace the mean profile $\mu$ with \autoref{eq:ring_profile} and include $\alpha_{\rm in, out},~r_0$ as parameters to be estimated as part of the HIBI posterior. Compared to \autoref{eq:image_model}, we kept the other parameters and priors identical, except we increased the number of pixels in the raster to $95\times95$ to better resolve the mean image structure when $\alpha_{\rm}$ was large. Specifically, we maintain the same priors for $\bm{r}$ and the related its related parameters $\rho$ and $\sigma$. The only remaining piece is to specify the priors for the ring parameters. For the characteristic radius $r_0$, we used a uniform prior $\mathcal{U}(5\,\muas, 40\,\muas)$; for the inner power law, we used the uniform prior $\mathcal{U}(0, 10)$, and for the outer power law, we used the prior $\mathcal{U}(1, 10)$. Setting $\alpha_{\rm in} = 0$ yields profiles without central depression, i.e., disk-like profiles. 

Note that while in this section we are mainly concerned with the parameters $r_0, \alpha_{\rm in, out}$, we will also produce a new set of images of \m87. Given that \m87 has been established as a ring, we expect the uncertainty of the image to decrease, since a large portion of the image will be described by the simple profile $\mu_r$. One potential concern with this informative model for $\mu$ is that it can bias the results if there is no ring in the image. To test this, we repeated the synthetic data tests from \autoref{sec:synthetic_data}. Generally, we found better results with the ring prior than the previous Gaussian prior image. We suspect that the reason for this is that the ring mean image better describes the general image structure for 4 out of 5 of the synthetic data sets. The profile \autoref{eq:ring_profile} is a near-perfect match for the symmetric ring and disk models, and the GMRF models the brightness asymmetry in the crescent and GRMHD models. The only nontrivial result is the double Gaussian blob model. In this case, the ring radius matches the blob separation, and the GMRF then modulates the rest of the brightness to match the true on-sky image. For more details on the results, see \autoref{appendix:ring_synthetic}.

Moving to the estimation of \m87's ring parameters, we repeat the analysis of \autoref{sec:M87} using this ring mean image prior. To fit the \m87 data, we use the same sampling procedure as \autoref{sec:inference}. The image reconstructions are shown in \autoref{fig:m87ring}. Qualitatively, the ring morphology appears similar to \autoref{fig:map_v_post}, with a circular ring close to $\sim 40 \muas$ and brighter at the bottom. To extract the ring parameters, we again use \vida with the same fourth-order ring template. The parameter estimates relative to the more agnostic prior are shown in \autoref{fig:ring_params}. We find consistent values with the Gaussian profile. However, the uncertainty of the diameter and width of the ring is reduced by around $50\%$, and the brightness asymmetry and position angle by $10-20\%$. This reduced uncertainty is expected, as the observed ring more closely matches the assumed $\mu$ structure, thereby reducing the variance of the GMRF and lowering the overall uncertainty in the image reconstructions. 

Comparing both the ring and Gaussian profile HIBI imaging results to the original EHT results \citepalias{M87EHTCIV}, we find consistent results, although the uncertainties reported tend to differ. For both the Gaussian and ring profiles, the measured uncertainties for the asymmetry and position angle reported by HIBI tend to be larger than those originally reported by the EHT. This is likely because the uncertainty reported by the EHT did not account for the uncertainty in image reconstruction given a set of hyperparameters. Instead, the EHT's hyperparameter surveys focused on estimating changes in the image structure due to different hyperparameter choices. HIBI estimates both the image and hyperparameter uncertainties jointly. For the ring width, we find that both the Gaussian and ring profiles give consistent measurements, unlike the EHT results, which only reported an upper bound of $20\,\muas$.

To test whether HIBI's ring width estimate is reliable, we imaged synthetic data of a symmetric delta ring blurred with a Gaussian with FWHM of $5\,\muas$, $10\,\muas$, $15\,\muas$, and $20\,\muas$ using both the Gaussian and ring profiles. The detailed results are shown in \autoref{app:ring_thickness}. We found that the Gaussian profile recovered the correct width for the $10, 15, 20\,\muas$ cases. However, for the $5\,\muas$ synthetic data, the Gaussian profile reconstruction was biased high, providing a measurement of $8.0^{+0.7}_{-0.8}\,\muas$, rather than the actual value of $5.7\,\muas$ measured by \vida. However, the ring profile $\mu_r$ accurately measured the ring width for all cases, likely due to the ring profile being a better match to the true image. Given that the synthetic data tests demonstrate the reliability of the measurement of the width of the ring profile, we believe that it is an accurate measurement of the width of \m87 in 2017. Combining the measurements of \m87 across all 4 days and 2 frequency bands, the total width estimate is $9.3 \pm 1.3\, \muas$. 

\begin{table}[!h]
    \centering
    \caption{Ring profile estimates LO band (68\% CI)}
    \vspace{-4mm}
    \begin{tabular}{l|cccc}
    \hline\hline
     Param.& April 5 & April 6 & April 10 & April 11  \\
    \hline
     $2r_{\rm peak}$\,$(\muas)$ & $34.4^{+1.7}_{-1.8}$ & $34.6^{+1.5}_{-1.6}$ & $36.0^{+2.0}_{-1.8}$ & $36.4^{+1.7}_{-1.5}$\\
     $2 r_0$ $(\muas)$         & $32.6^{+1.4}_{-1.6}$ & $32.8^{+1.2}_{-1.5}$ & $34.2^{+1.8}_{-1.7}$ & $34.6^{+1.4}_{-1.3}$\\
     $\alpha_{\rm in}$         & $58.8^{+27.6}_{-29.0}$ & $62.8^{+23.9}_{-31.8}$ & $55.4^{+30.5}_{-31.5}$ & $54.7^{+31.5}_{-30.9}$\\
     $\alpha_{\rm out}$        & $1.7^{+0.4}_{-0.3}$ & $1.6^{+0.4}_{-0.3}$ & $2.3^{+0.8}_{-0.6}$ & $2.5^{+0.6}_{-0.5}$\\
     \hline
     {\small \vida est.}\\
     \hline
    $d$ $(\muas)$ & $41.9^{+1.1}_{-1.1}$ & $42.4^{+1.0}_{-1.2}$ & $43.2^{+1.3}_{-1.1}$ & $43.4^{+1.0}_{-0.9}$ \\
    $w$ $(\muas)$ & $8.9^{+1.1}_{-1.2}$ & $9.0^{+1.1}_{-0.9}$ & $9.1^{+1.6}_{-1.2}$ & $9.1^{+0.8}_{-1.1}$\\
    $A$           & $0.29^{+0.04}_{-0.04}$ & $0.27^{+0.05}_{-0.04}$ & $0.29^{+0.05}_{-0.05}$ & $0.25^{+0.03}_{-0.04}$ \\
    $\xi$ $(\deg)$ & $160.2^{+6.6}_{-9.9}$ & $159.0^{+7.6}_{-10.5}$ & $177.9^{+6.1}_{-5.5}$ & $176.3^{+4.7}_{-6.6}$ \\
     \end{tabular}
    \label{tab:ring_profile_params}
\end{table}

Turning to the estimation of the parameters of the ring profile, \autoref{tab:ring_profile_params} reports the parameter estimates for the peak radius, characteristic radius, and the inner and outer power-law slope. Interestingly, we find that the inner slope is quite large, typically ranging from $10$ to $40$, while the outer power law slope is closer to between $ 1.5$ and $2.5$ for all four days. This suggests that the ring emission tends to peak closer to the inner edge of the shadow and declines more gradually.

 \section{Application to VLBA Data}\label{sec:AGN}

 \begin{figure*}[!ht]
    \centering
    \includegraphics[width=1.0\linewidth]{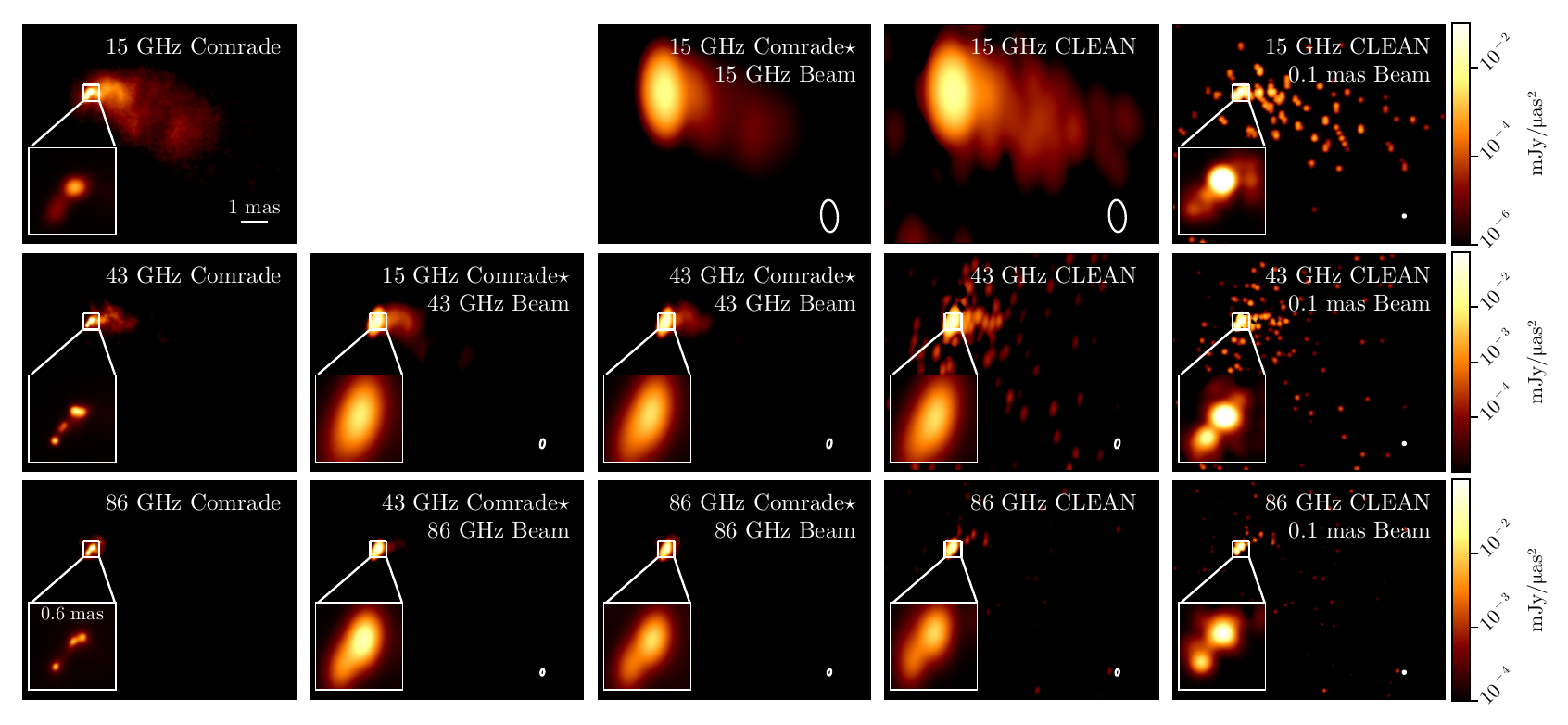}
    \caption{The mean \comrade reconstructions of \oj287 on Nov 1 and 2 2016 at 15\,GHz, 43\,GHz, and 86\,GHz compared to CLEAN images. The first(leftmost) column shows the \comrade reconstructions results at its native resolution. The second column shows the \comrade reconstructions blurred to match the beam at the next higher frequency. For example, the second row is the 15\,GHz reconstruction blurred to match the resolution of the 43\,GHz CLEAN reconstructions. The third column shows the \comrade reconstructions blurred to match the resolution of the native CLEAN reconstructions shown in the fourth column. Overall, we find that \comrade's super resolution of the core is supported by the image reconstructions at higher frequencies.}
    \label{fig:agn}
\end{figure*}

Although the algorithms developed in this paper focus on black hole imaging with the EHT, they apply to any VLBI array. To demonstrate HIBI's capabilities, we now consider VLBA AGN data. Specifically, we analyzed the AGN \oj287 observations on November 1 at 43 and 86\,GHz and on November 2 at 15 GHz. For the 43 and 86\,GHz data, we utilize public data from the Boston University BEAM-ME project \citep{bu2016, bu2017, bu2022}. For the 15\,GHz data, we use public data from the MOJAVE AGN monitoring program \citep{mojave}.

For the 15 GHz data, a $144\times112$ raster with a FOV of $7.8\times6$\,mas was chosen. For the 43 GHz data, a $192\times96$ raster with a $4\times2$\,mas raster was used. For the 86\,GHz data, a $96\times96$ raster with a FOV of $1\times1$\,mas was used. The raster size was chosen for efficiency when computing the NFFT. For the total flux of each image, we use the total flux reported from each experiment: $F_{{\rm 230}} = 7.17\,{\rm Jy}$, $F_{\rm 86} = 6.50\,{\rm Jy}$, and $F_{86 \rm GHz} = 4.58\,{\rm Jy}$. Similarly to the above, a first-order GMRF prior is assumed for each dataset, with the same priors on the hyperparameters as mentioned in \autoref{sec:M87}. For the a priori image structure of the GMRF, we used
\begin{equation}\label{eq:muoj}
    \mu_{\rm {OJ}}(\bm{x} | f_b) = 
        (1-f_b)\mu_{\rm core}(\bm{x}) + f_b/K.
\end{equation}
Here $\mu_{\rm core}(\bm{x})$ is a Gaussian profile with FWHM $400\,\muas$, $f_b$ is the fraction of the flux in the background component, and $K$ is the total number of image pixels. This profile corresponds to a core image with a potential constant background whose fractional flux, $f_b$, is also a free parameter fit during sampling. The point of the Gaussian core component is to encourage \oj287's core to reside near the phase center of the image, which the data does not constrain because we refit for instrumental gain phases as mentioned below. Finally, we used the exponential map \autoref{eq:delta_exp} to map from $r_{ij}$ to $\delta_{ij}$ due to the potentially high-dynamic range image structure.

For simplicity, we used the publicly available calibrated data. However, we refit the gain solutions to correct for any residual calibration errors. Given that we used well-calibrated data, we averaged each dataset over a ``scan'' using the \texttt{scan\_average} function from \texttt{Pyehtim.jl} and added 0.2\% systematic error. Similarly to the \m87 analysis, we fit the gain phases for each station using a von Mises prior with the concentration parameter $\pi^{-2}$. We used a normal distribution for the log-gain amplitudes with a mean of zero and a standard deviation of $0.25$ for all stations. The fitting and sampling strategy is identical to \autoref{sec:inference}. 

\autoref{fig:agn} shows the resulting fits. The first column shows the native resolution image reconstruction results at 15, 43, and 86 GHz from top to bottom. For all frequencies, a qualitatively similar core structure consisting of 2 to 3 point-like objects was recovered using HIBI. The third column shows the image reconstructions blurred to match the CLEAN images (fourth column), demonstrating mostly consistent results, although the HIBI images from \comrade images are less noisy. The second column of the figure shows the 15(43)\,GHz HIBI reconstruction, blurred to match the 43(86)\,GHz CLEAN resolution in the second(third) rows. For all three frequencies, we see that the HIBI reconstruction achieves significantly improved resolution compared to the CLEAN reconstruction blurred with the nominal beam. The improved resolution observed at 15 and 43\,GHz is similar to the observed structure at higher frequencies and supports the conclusion that the improved resolution is faithful to the true source structure. Note, however, that the improved resolution is significantly higher near the core region of the image, where we have the largest effective SNR of the image. For the more extended emission observed at 15\,GHz, the resolution is more similar to CLEAN reconstructions. The result demonstrates how effective resolution depends on the SNR of various image components \citep{Lobanov}.

\section{Summary}\label{sec:summary}

This paper presents a new perspective on VLBI imaging, casting it as a hierarchical Bayesian inference problem. HIBI separates imaging into two distinct types of parameters: the latent parameters, which are the image itself, and the higher-level or structural parameters, sometimes referred to as hyperparameters. Within this hierarchical framework, HIBI can reconstruct both types of parameters and estimate their joint uncertainty.

To demonstrate HIBI, we implemented a scalable version using Markov random fields and demonstrated how physical information can be incorporated, e.g., ring-like morphology, into the model. The HIBI algorithm itself is available in the \comrade VLBI imaging software, and examples can be found in the online documentation\footnote{\url{https://ptiede.github.io/Comrade.jl/stable/}}.

To test the applicability of HIBI to various image structures, we repeated the synthetic data tests from \citetalias{M87EHTCIV}. In all cases, after marginalizing over the gain and hyperparameter parameters, we found that HIBI reconstructs the correct image morphology and quantitative features for EHT 2017 coverage. We then applied HIBI to the 2017 EHT data, reproducing the original EHT results, and measuring \m87's diameter, width, brightness asymmetry, and position angle. 

While the HIBI posterior can be used to construct images marginalized over different structural and instrumental assumptions, we can also marginalize over image fluctuations to estimate more fundamental parameters. In \autoref{sec:ring} we demonstrated how we can use HIBI to directly measure properties of the ring, such as its radius and profile shape, after marginalizing over different image fluctuations. We anticipate that this approach to feature extraction will be highly beneficial for science cases where the rough morphology or physics is known a priori. For example, the demographics of marginally resolved black hole shadow candidates from future missions like the ngEHT \citep{ngeht,Doeleman_2023} and the Black Hole Explorer \citep{BHEX}, could utilize the ring modeling in this paper to measure the radius of the shadow of other black holes. This approach to imaging could also be expanded to include more physically interesting models, including semi-analytic accretion models \citep[e.g.,][]{RIAF, Palumbo_2022, BayesPhotogram} where the imaging component would focus on modeling the stochastic turbulence in plasmas around black holes. 

In \autoref{sec:AGN} we applied HIBI to non-EHT data, imaging the well-known blazar \oj287 and demonstrating the reliability of HIBI's superresolution by comparing the reconstructions across frequency for near-simultaneous observations. We found that HIBI dramatically improved the image reconstruction resolution in the bright core region, and the core morphology was confirmed upon examining higher-frequency data.

\begin{acknowledgments}
The authors thank Sara Issaoun, Iniyan Natarajan, Aleksander (Sasha) Plavin, and Andrew Chael for informative discussions during paper writing. This publication was supported by the AWS Impact Computing Project at the Harvard Data Science Initiative, Award \#A61166. We acknowledge financial support from the National Science Foundation (AST-2307887). This publication is funded in part by the Gordon and Betty Moore Foundation, Grant GBMF12987. This work was supported by the Black Hole Initiative, which is funded by grants from the John Templeton Foundation (Grant \#62286) and the Gordon and Betty Moore Foundation (Grant GBMF-8273) - although the opinions expressed in this work are those of the author(s) and do not necessarily reflect the views of these Foundations. This research has made use of data from the MOJAVE database that is maintained by the MOJAVE team \citep{mojave}. This study makes use of VLBA data from the VLBA-BU Blazar Monitoring Program (BEAM-ME and VLBA-BU-BLAZAR;
\url{http://www.bu.edu/blazars/BEAM-ME.html}), funded by NASA through the Fermi Guest Investigator Program. The VLBA is an instrument of the National Radio Astronomy Observatory. The National Radio Astronomy Observatory is a facility of the National Science Foundation operated by Associated Universities, Inc.
\end{acknowledgments}

\vspace{5mm}
\facilities{EHT, (ALMA, APEX, IRAM, JCMT, LMT, SMA, SMT, SPT), VLBA}


\software{\comrade \citep{comrade}, \texttt{AdvancedHMC.jl} \citep{xu2020advancedhmc}, \ehtim \citep{Chael_2016, Chael_2018}, \texttt{Enzyme} \citep{NEURIPS2020_9332c513, 10.1145/3458817.3476165, 10.5555/3571885.3571964}, Julia \citep{julia}, \texttt{NFFT.jl} \citep{NFFTjl}, \texttt{Makie.jl} \citep{DanischKrumbiegel2021}, \texttt{Optimization.jl} \citep{dixit_optimizationjl_2023}, \texttt{Optimisers.jl} (\url{https://github.com/FluxML/Optimisers.jl})}


\newpage
\appendix



\section{Imaging Parameters}
\autoref{tab:imaging_params} shows the parameters used for the image reconstructions for the various sets of data considered in this paper.

\begin{table*}[!h]
    \caption{Imaging Parameters}
    \begin{tabular}{l|ccccc}
    \hline\hline
     Data & $\mu(x, y)$ & $\eta(r)$ & GMRF Order & Raster size & FOV \\
     \hline
     EHT \m87/synth. (Gauss.) & $\mu_g$ (\ref{eq:mug})            & $\eta_{\rm sp}$ (\ref{eq:softplus}) & 1 & $64\times64$ & ($200\,\muas, 200\muas$)\\
     EHT \m87/synth. (ring)     & $\mu_r$ (\ref{eq:ring_profile}) & $\eta_{\rm sp}$ (\ref{eq:softplus}) & 1 & $95\times95$ & ($200\,\muas, 200\muas$) \\
     VLBA OJ\,287 (15\,GHz)     & $\mu_{\rm OJ}$ (\ref{eq:muoj})  & $\eta_{\rm sm}$ (\ref{eq:delta_exp})& 1 & $143\times111$ & ($7.8\,{\rm mas}, 6\,{\rm mas}$) \\
     VLBA OJ\,287 (43\,GHz)     & $\mu_{\rm OJ}$ (\ref{eq:muoj})  & $\eta_{\rm sm}$ (\ref{eq:delta_exp})& 1 & $191\times95$ & ($4\,{\rm mas}, 2\,{\rm mas}$) \\
     VLBA OJ\,287 (86\,GHz)     & $\mu_{\rm OJ}$ (\ref{eq:muoj})  & $\eta_{\rm sm}$ (\ref{eq:delta_exp})& 1 & $95\times95$ & ($1\,{\rm mas}, 1\,{\rm mas}$) \\
    \end{tabular}
    \label{tab:imaging_params}
\end{table*}

\section{Impact of Correlation in the Image Prior}
To explore the impact of including correlation in the image prior, this section will re-image the April 11 lo-band data using two commonly used priors in Bayesian imaging from \citet{BroderickHybrid} and \citet{dmc}. The \citet{BroderickHybrid} inspired prior assumes a log-uniform prior for the pixel fluxes $\bm{F}$ with lower bound $A$ and upper bound $B$. This prior tends to enforce a high degree of sparsity in the data and does not have any hyperparameters other than the number of pixels in the raster and the field of view of the image. The \citet{dmc} prior, assumes that $\bm{F}$ is drawn from a symmetric Dirichlet distribution with probability density
\begin{equation}
    p_{\mathcal{D}}(\bm{F} | \alpha) = \frac{\Gamma(c K)}{\Gamma(c)^K}\prod_{I=1}^K F_{I}^{c - 1},
\end{equation}
where $c$ is the concentration parameter and enforces a global notion of smoothness or sparsity. For $c = 1$, this distribution corresponds to the uniform distribution on the simplex. For $c < 1$, the prior tends to promote sparsity in the image, while for $c > 1$, it tends to prefer homogeneous images. Given that it is unclear what to choose for $c$ a priori, we included it as a hyperparameter during imaging. Similarly to the GMRF prior, we used a hyperprior that tends to prefer simple images, i.e., images with similar flux everywhere, i.e., $c > 1$. To enforce the preference for a smooth image, we use an inverse gamma prior with $\alpha = 6,\, \beta = 9.25$. This distribution has its mode at $c \approx 1.3$ and has $10\%$ of its probability mass $c < 1.0$.

\begin{figure*}[!t]
    \centering
    \includegraphics[width=\linewidth]{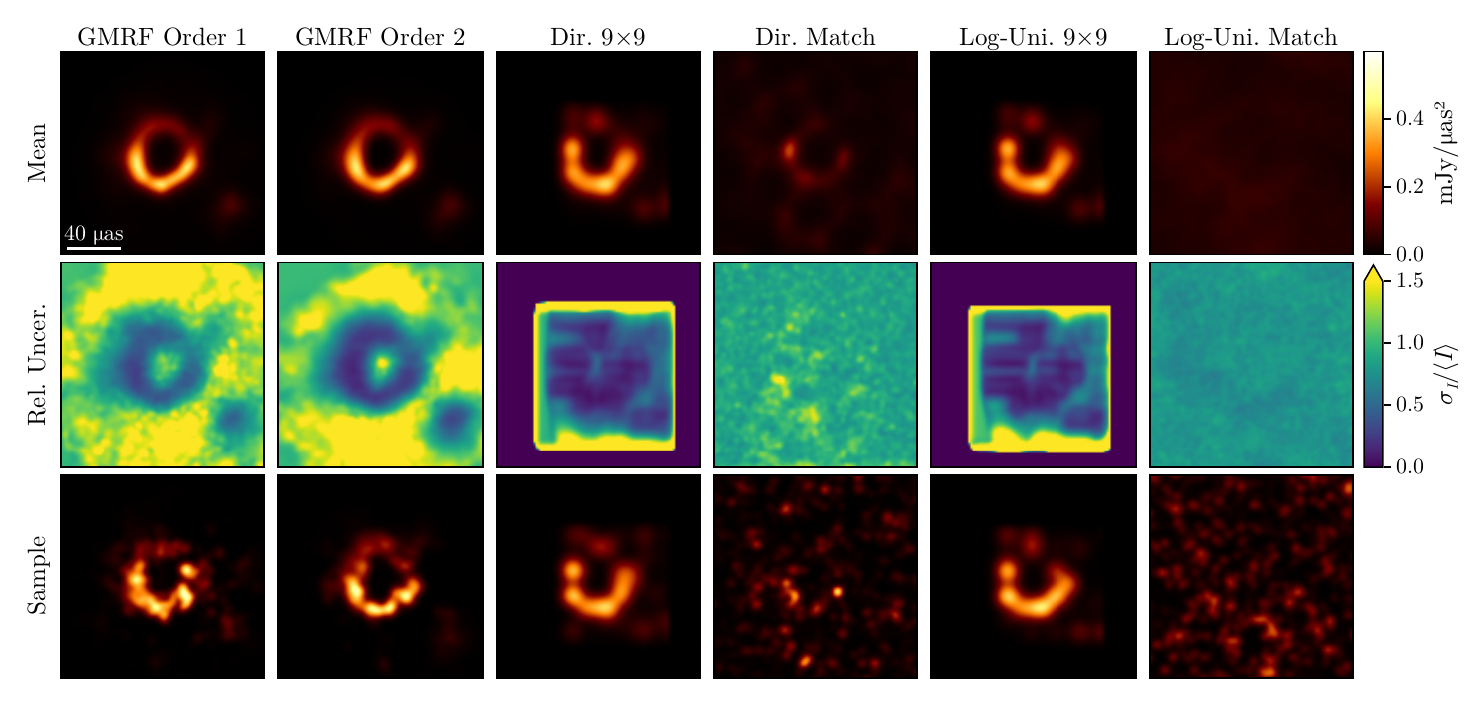}
    \caption{Comparison between three alternative Bayesian imaging priors. }
    \label{fig:uncorr}
\end{figure*}

\begin{table}[!h]
    \centering
    \caption{April 6 Ring Parameters}
    \begin{tabular}{l|cccc}
        \hline\hline
        Prior       &  $d\; (\muas)$ & $w\; (\muas)$ & $A$ & $\xi (^\circ)$ \\
        \hline        
        GMRF Order 1 $(\mu_g)$       &  $41.0^{+2.3}_{-3.5}$ & $11.1^{+2.0}_{-2.5}$ &  $0.27^{+0.09}_{-0.06}$ & $155.2^{+14.6}_{-19.5}$ \\
        GMRF Order 2 $(\mu_g)$       & $41.4^{+2.2}_{-3.1}$ & $11.5^{+1.9}_{-1.6}$ & $0.26^{+0.07}_{-0.05}$ & $158.0^{+11.6}_{-20.9}$ \\
        Dirichlet ($9\times9$)   &  $40.3^{+1.4}_{-1.2}$ &  $14.8^{+1.0}_{-0.8}$ & $0.29^{+0.04}_{-0.04}$ & $177.3^{+3.7}_{-4.4}$  \\
        Dirichlet ($64\times 64$)   &  $36.7^{+9.8}_{-9.5}$ & $7.6^{+5.8}_{-2.4}$ & $0.33^{+0.17}_{-0.16}$ & $128.4^{+39.2}_{-44.7}$ \\
        Log-Uniform ($9\times9$) &  $40.5^{+1.2}_{-1.1}$ & $14.2^{+0.9}_{-0.5}$ & $0.31^{+0.04}_{-0.05}$ & $176.9^{+4.4}_{-3.5}$\\
        Log-Uniform ($64\times64$)   &  $53.4^{+6.6}_{-15.2}$ &$8.2^{+27.1}_{-3.4}$ & $0.33^{+0.17}_{-0.26}$ & $157.0^{+146.6}_{-94.9}$  \\
    \end{tabular}
    \label{tab:EHT_ring_parameters}
\end{table}

As the log-uniform and Dirichlet priors did not include correlation parameters, we ideally should search over different FOVs and raster dimensions to find the optimal number of pixels according to some cost function, such as the Bayesian evidence or Bayesian leave-one-out cross-validation. However, in this paper, we instead fixed the values of the field of view of to $80\muas$ since it was large enough to contain all the flux seen in \autoref{fig:map_v_post}. To select the pixel size, we followed \citet{BroderickHybrid} and used a pixel size of $\sim 12\muas$, which corresponds to a $7\times 7$ raster. This is far below the resolution of the GMRF prior in the main text, so we also considered a raster identical to the GMRF run to demonstrate the impact of enforcing zero correlation in the image. To sample the image posterior, we employed the same procedure from \autoref{sec:inference}. 

The image reconstructions for the log-uniform (left), Dirichlet (middle), and GMRF (right) priors are shown in \autoref{fig:uncorr}. Overall, the qualitative image structure looked similar except for the uncorrelated priors that use a large FOV and a large number of pixels. Both lower resolution and log-uniform priors produce similar rings in appearance, and qualitatively, as can be seen in \autoref{tab:EHT_ring_parameters}. However, when the FOV and number of pixels were increased to match the GMRF values, both the Dirichlet and log-uniform prior struggled to identify a unique image structure. For the Dirichlet prior, we see that the mean image displays a ring-like feature, albeit at a low significance. For the log-uniform prior, no significant ring-like feature can be identified in the image reconstruction, and samples from the posterior look like random noise. This random noise is due to the extreme overfitting. Since no spatial correlation is enforced, the number of degrees of freedom ($64\times64$) in the image is much greater than the size of the data ($\sim 200$ visibilities). As a result, the data are not very informative relative to the model, and the posterior starts to approach the prior distribution. If we view this model within the HIBI framework, the number of pixels could be included as a discrete hyperparameter in the model. \citet{BroderickHybrid} approximately demonstrated that in such a scheme, high-resolution rasters are heavily disfavored compared to lower resolution ones.

\section{EHT \& GMRF Prior Sensitivity to On-Sky Ring Thickness}\label{app:ring_thickness}
 \begin{figure}[!h]
     \centering
     \includegraphics[width=0.50\linewidth]{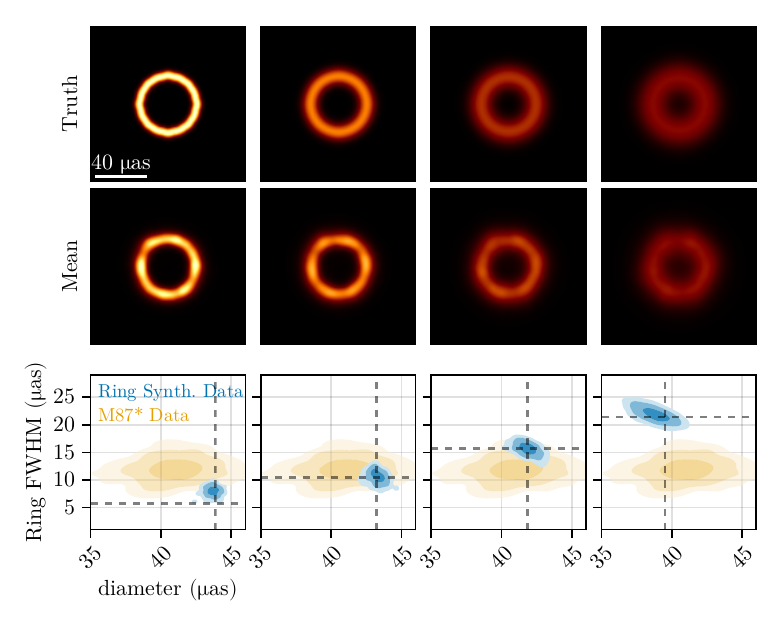}
     \hspace{-4mm}
     \includegraphics[width=0.50\linewidth]{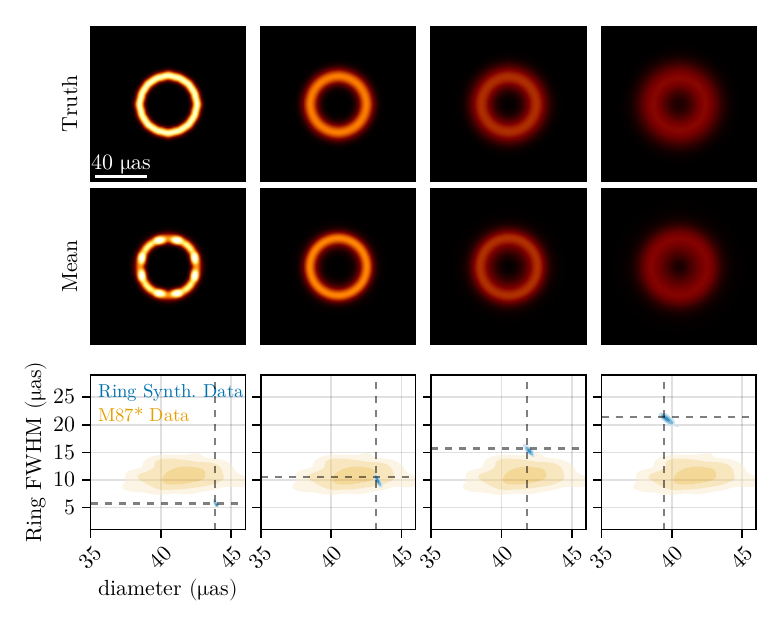}
     \caption{Testing the sensitivity of the EHT array to different ring thicknesses for the Gaussian (left group) and ring (right group) mean prior. For each group, the columns show rings with intrinsic widths $5\muas$, $10\muas$, $15\muas$, and $20\muas$ from left to right. The first row is the ground truth, the middle row is the mean reconstruction using the first-order GMRF, and the bottom row shows the posterior density of the ring width vs. diameter, with contours denoting the $50\%$, $90\%$, and $95\%$ credible intervals. The measured \m87 width and diameter averaged over the observing days and frequency are shown by the orange region.}
     \label{fig:ringwidth}
 \end{figure}

 The thickness of \m87's ring was reported by the EHT as an upper limit of $20\,\muas$ across imaging and modeling methods \citep{M87EHTCIV,M87EHTCVI}. Later work \citep{LockhartGrallaI, LockhartGrallaII} analyzed the 2017 EHT data with a variety of geometric models and found that the fractional width of the ring model was $\leq 0.25$, which for a $\sim 40\,\muas$ ring, gives a width of $\sim 10\,\muas$. However, the authors noted that the fractional width bound could increase to $0.3-0.4$ or $12-16\,\muas$ when using a different approximate closure likelihood. 

Using \comrade and the GMRF prior with a Gaussian profile, we measured the thickness of the ring to be $11.7 \pm 0.7\,\muas$ averaged over the four days and two frequency bands. Meanwhile, for the ring profile, we found a similar width of $11.0\pm0.5\,\muas$. Both of the measurements are consistent with \citet{M87EHTCIV}, \citet{M87EHTCVI}, and \citet{LockhartGrallaII}. However, due to the finite resolution of the EHT, the width measurements may be biased. Namely, the EHT isn't sensitive to changes in the ring width below a certain limit. To test whether HIBI's ring thickness measurement is reliable for the scales measured by EHT, we imaged rings with several different thicknesses. Namely, we used the same azimuthally symmetric delta ring from \autoref{sec:synthetic_data}, but blurred using different Gaussian kernels with FWHM of $5, 10, 15, 20 \muas$. We then imaged the said synthetic data using both the Gaussian and ring profiles.  

The results are shown in \autoref{fig:ringwidth}. We found that HIBI recovers the true diameter and width of the rings for the 10, 15, and 20\,$\muas$ rings for both $\mu_g$ and $\mu_r$. However, for the 5\,$\muas$ ring, \autoref{fig:ringwidth} the Gaussian profile reconstructions have widths that are slightly too large compared to the ground truth measurement $5.1\,\muas - 10\,\muas$. However, the ring profile $\mu_r$ recovers the true width of the ring. To ensure that $\mu_g$ was not biased by the rasterization resolution, we reran the $5\,\muas$ synthetic data with a higher-resolution raster and found equivalent results. Therefore, this bias is likely due to a combination of the EHT's limited resolution and the fact that the Gaussian profile does not accurately match the true value. 

For \m87 we found that both the ring and Gaussian profile provided identical width measurements. Additionally, \m87's measured width value more closely matches the $10\,\muas$ ring data, which was recovered for both profiles considered. Therefore, we believe that HIBI's estimate of the ring width of \m87 is reliable.

\section{Robustness of the A Priori Ring Profile to Different Structures}\label{appendix:ring_synthetic}
 \begin{figure*}[!h]
     \centering
     \includegraphics[width=\linewidth]{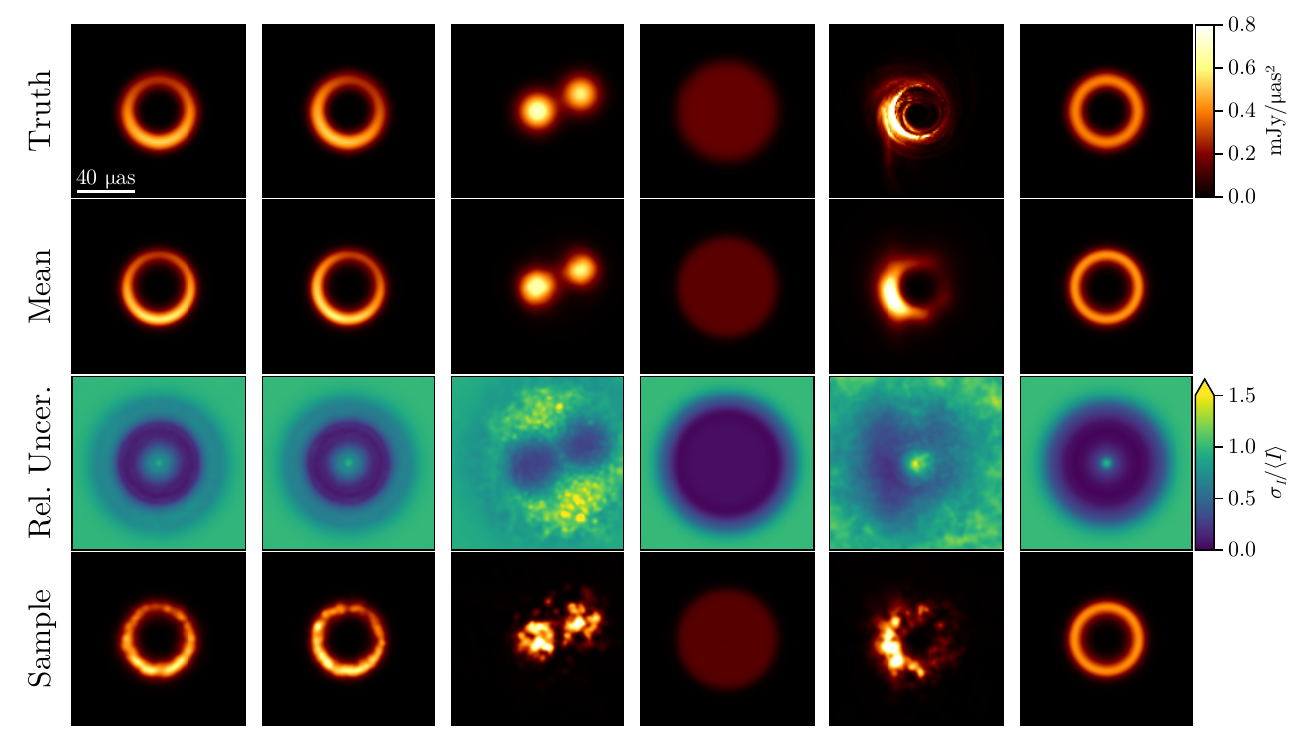}
     \caption{Reconstructions of synthetic data from \autoref{sec:synthetic_data}, but using a ring prior on the mean image of the MRF.}
     \label{fig:ringfiltering}
 \end{figure*}

In this section, we demonstrate that the EHT data is informative enough to produce non-ring images even when a ring model is used for $\mu(x,y)$. This is complementary to the synthetic data tests in \citetalias{M87EHTCIV}, which demonstrate that the EHT can reconstruct various image structures with minimal assumptions about the source morphology. 

 For the data, we used the same synthetic data described in \autoref{sec:synthetic_data}. However, for the a priori profile $\mu(x,y)$, we used the ring profile \autoref{sec:ring}. This profile consists of a ring, although the ring radius and the inner and outer power-law slopes are included as parameters in the model. Furthermore, the inner power-law prior has support around zero, meaning that a flat disk is in the space of prior images.

 The imaging results are shown in \autoref{fig:ringfiltering}. For the ring synthetic data, we observed that the ring prior significantly reduces the uncertainty in the image reconstructions, because the profile $\mu$ is a near-perfect model of the image. The disk reconstruction also improved considerably as it is included in the family of $\mu(x,y)$ profiles. Encouragingly, even when a ring-like structure was assumed, the double synthetic data were constructed correctly. This result implies that the EHT data are informative enough to create non-ring images even when a ring prior is assumed when using HIBI. Therefore, we conclude that the ring profile does not rigidly enforce ring-like images when analyzing the 2017 EHT data.

 \section{Robustness of \m87 image to Different Gain Assumptions}\label{app:closures}
  \begin{figure}[!h]
     \centering
     \includegraphics[width=\linewidth]{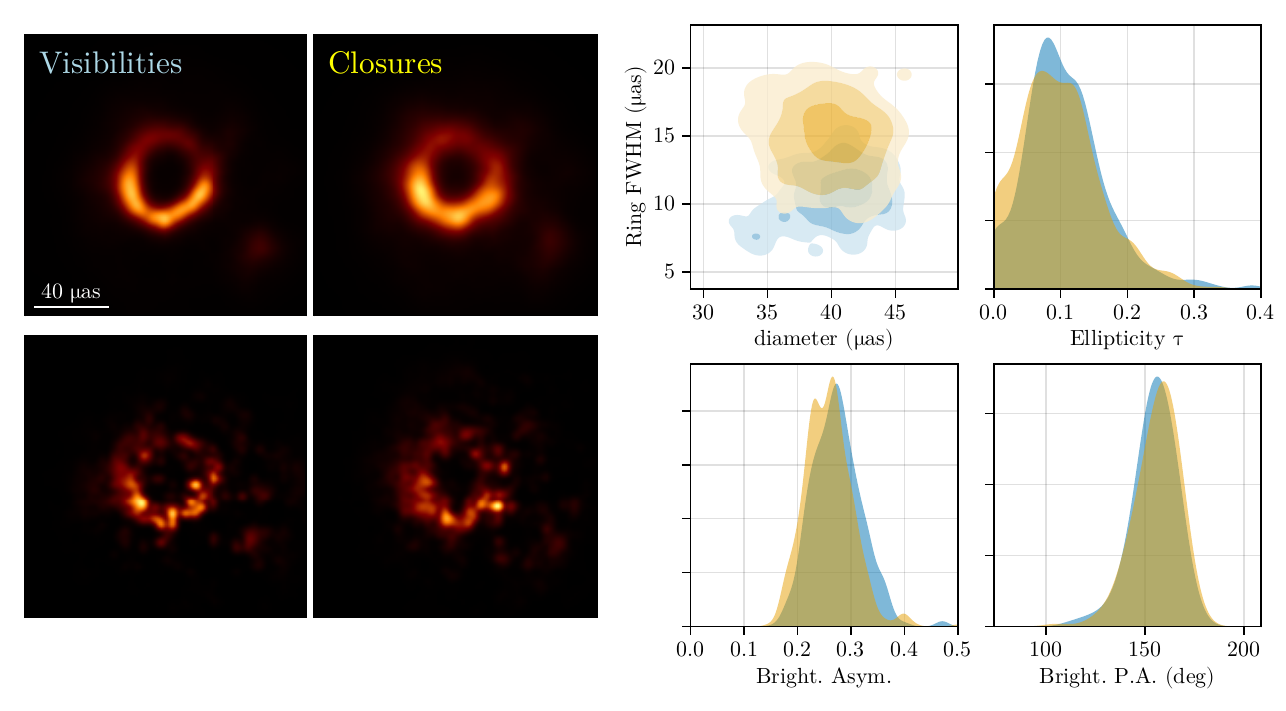}
     \caption{Comparison between complex visibility (blue) and closure only (orange) reconstructions using the April 6 LO band data. The top row of the images shows the mean image, while the bottom are random draws from their respective posterior. The right grid of images show the parameter estimates computed using the VIDA template matching approach described in \autoref{sec:synthetic_data}}
     \label{fig:closures}
 \end{figure}

\subsection{Closure Likelihood}
Fitting complex visibilities directly mean that we require an instrument model to describe both atmospheric and instrumental effects. Since VLBI imaging is typically only interested in the image posterior $p(\bm{\theta}| \bm{V})$, another mathematically equivalent approach is to first marginalize over the instrument model parameters $\bm{s}$:
\begin{equation}
    p(\bm{\theta} | \bm{V}) = \frac{p(\bm{\theta})}{p(\bm{V})}\int p(\bm{V} | \bm{\theta}, \bm{s})p(\bm{s})\dd\bm{s}.
\end{equation}

\citet{BlackburnClosures} demonstrated that the marginal image posterior is equivalent to fitting closure products in the high SNR limit when improper flat priors are assumed for the instrumental gains. This assumption can be viewed as a maximally conservative view of our a priori knowledge of the instrument. Therefore, to explore the impact of our instrument model in \autoref{sec:M87}, this section considers image reconstructions using log-closure amplitudes and closure phases. 

Closure phases are defined as the argument of the bispectrum of three measured visibilities:
\begin{equation}
    \psi_{ijk} = \arg V_{ij}V_{jk}V_{ki},
\end{equation}
and provide the gain-invariant phase information of the set of measured visibilities. Amplitude information can be recovered by analyzing the closure amplitudes:
\begin{equation}
    A_{ijkl} = \frac{|V_{ij}||V_{kl}|}{|V_{ik}||V_{jl}|}.
\end{equation}
Instead, we consider the log-closure amplitudes $\mathcal{A} = \log A$. The reason is that log-closure amplitudes are related to log amplitudes by a linear transformation, and their errors are better approximated by Gaussian noise than the direct closure amplitudes \citep{TMS, Themis}.

For high-SNR measurements ($\gtrsim 3$), the closure likelihood is approximately a correlated Gaussian \citep{TMS, Themis, BlackburnClosures}. For the closure phases, we used the approximate likelihood
\begin{equation}
    \mathcal{L}(\bm{\psi} | \bm{\theta}) \propto
            \exp\left(-\frac{1}{2}\Delta\bm{c}^\dagger\Sigma_{\psi}\Delta\bm{c}\right),
\end{equation}
where $\Delta\bm{c}_I = e^{i\psi_I} - e^{i\psi_{\theta,I}}$, and $\psi_{\theta, I}$ is the image model, $I$-th closure phase. Note that the normalization of this likelihood is not easily calculable; however, it amounts to a scaling constant, so it does not affect posterior estimation. 

Similarly, for the log closure amplitudes, we make a similar high-SNR Gaussian approximation for the likelihood:
\begin{equation}
    \mathcal{L}(\bm{\mathcal{A}} | \bm{\theta}) \propto 
            \exp\left(-\frac{1}{2}\Delta\bm{\mathcal{A}}^T\Sigma_{\mathcal{A}}\Delta\bm{\mathcal{A}}\right).
\end{equation}
Here $\Delta\mathcal{A}_I = \mathcal{A}_I - \mathcal{A}_{\theta, I}$, and $\mathcal{A}_{\theta, I}$ is the image model, $I$-th log-closure amplitude.

To construct the set of closure phases and log-closure amplitudes, we follow \citet{BlackburnClosures}. Let $L_{\mathcal{A}}$ and $L_{\psi}$ denote the linear maps from log amplitudes/phases to log-closure amplitudes/phases. We will assume that $L$ is constructed such that both are full-rank matrices. Then we have that 
\begin{equation}
    \Sigma_{\mathcal{A},\psi} = L_{\mathcal{A},\psi}\frac{\bm{\sigma}^2}{|\bm{V}|}L_{\mathcal{A},\psi}^T.
\end{equation}
The total closure likelihood is therefore given by
\begin{equation}\label{eq:closure_post}
    p(\bm{\theta} | \bm{V}) \approx 
    p(\bm{\theta} | \bm{\psi}, \bm{\mathcal{A}}) \propto
    \mathcal{L}(\bm{\psi} | \bm{\theta})\mathcal{L}(\bm{\mathcal{A}} | \bm{\theta}) p(\bm{\theta}).    
\end{equation}

 To compare the impact of our gain priors on the images, we compare the results in \autoref{sec:M87} with closure-only fits. To simplify the comparison, we only show the results for the April 11 LO band. However, we found similar results on other days. For closure-only reconstructions, we flag any closures whose ${\rm SNR} < 3$ to remove closures that violate our Gaussian likelihood assumption. Otherwise, we use identical fitting and sampling procedures. 
 
 The imaging results are shown in \autoref{fig:closures}. We measured similar diameters, ellipticities, brightness asymmetries ($A$), and position angles ($\xi$) for both complex visibilities and closures. The major difference in the reconstruction is the width or resolution of the ring. That is, the complex visibility fits appear sharper, and the inferred ring width posterior is shifted slightly downwards. The difference is likely explained by the loss of information when moving to closures, and the flagging of low SNR points, which would tend to remove longer baselines and the measurements near \m87's amplitude null \citepalias{M87EHTCVI}. Regardless, this result implies that the measurements of \m87's properties in the main text are not significantly biased by the assumed instrument model.

\section{Eigenvalues of the First Order GMRF}\label{app:eigenvalues}
In this section, we prove \autoref{eq:det}.
The precision matrix $Q_1$ for the unit variance first order GMRF is 
\begin{equation}
    \bm{Q}_1 = \frac{1}{\rho^2}\mathds{1} + \bm{G}.
\end{equation}
To compute its eigenvalues, we first note that $\bm{G}$ is given by the Kronecker sum
\begin{equation}
    \bm{G} = \bm{D}_x\otimes \mathds{1} + \mathds{1}\otimes \bm{D}_y,
\end{equation}
where $\bm{D}$ is the tridiagonal matrix,
\begin{equation}
    \bm{D}_{ij} = \begin{cases}
        2 & i = j, \\
        -1 & i = j\pm1, \\
        0 & {\rm otherwise},
    \end{cases}
\end{equation}
the subscripts denote the number of pixels in the $x$ and $y$ directions, and $\otimes$ denotes the tensor product.

Given the form of $G$ its eigenvalues are just the sum of the eigenvalues of $D_x$ and $D_y$, i.e., the eigenvalues of $G$ are given by
\begin{equation}
    \lambda^G_{nm} = \lambda_{n} + \lambda_{m} \qquad 1 \leq n \leq N_x\;, 1 \leq m \leq N_y.
\end{equation}
Therefore, to find the spectrum of $G$, we need to find the eigenvalues for $D$.
First, note that the eigenvalues of the $N\times N$ matrix
\begin{equation}\label{eq:chebymat}
\begin{pmatrix}
     0  & -1 &  &  &  \\
    -1 &  \ddots & \ddots & & \\
       & \ddots & & &\\ 
\end{pmatrix},
\end{equation}
are given by 
\begin{equation}\label{eq:eigenvals}
    2\cos\left(\pi\frac{n}{N+1}\right), \qquad 1 \leq n \leq N.
\end{equation}
This equation can be derived by noting that the characteristic polynomial of \autoref{eq:chebymat} defines the recurrence relation,
\begin{equation}\label{eq:recur}
    T_{n}(\lambda) = \lambda T_{n-1}(\lambda) - T_{n-2}(\lambda),
\end{equation}
where $T_1 = \lambda$ and we set $T_{0} = 1$. Let $x = \lambda/2$, then \autoref{eq:recur} becomes
\begin{equation}
\begin{aligned}
    T_0(x) &= 1\\
    T_1(x) &= 2x\\
    T_{n}(x) &= 2x T_{n-1}(x) - T_{n-2}(x),
\end{aligned}
\end{equation}
which are the recurrence relations for Chebyshev polynomials of the second kind, whose roots give \autoref{eq:eigenvals}.

 Given \autoref{eq:eigenvals}, the eigenvalues for $D$ are given by
\begin{equation}
    \lambda_n = 2 + 2\cos\left(\pi\frac{n}{N+1}\right),
\end{equation}
and thus eigenvalues for $Q_1$ are
\begin{equation}
    \lambda^{Q_1}_{nm} = \rho^{-2} + 4 + 2\cos\left(\pi\frac{n}{N_x+1}\right) + 2\cos\left(\pi\frac{m}{N_y+1}\right),
\end{equation}
whose product gives \autoref{eq:det}.

Note that using a similar computation, one can show that the eigenvectors of $D$, are given by
\begin{equation}
    (\bm{v}_{n})_{j} \propto \sin\left(\pi\frac{nj}{N+1}\right),
\end{equation}
for the $j^{\rm th}$ component of the $n^{\rm th}$ eigenvector. This implies that the type 1 discrete sine transform diagonalizes the matrix $D$.

\section{Extending the GMRF priors}\label{appendix:higher-order}
While the GMRF prior used in the main text is flexible, it assumes that the noise structure of the image is conditionally dependent only on the nearest neighbors and that the noise structure is Gaussian. This appendix demonstrates how you can relax both constraints. 

\subsection{Non-Gaussian Markov Random Fields}
Recall that the general form of a Gaussian Markov Random Field (GMRF) is given by \autoref{eq:GMRF}. We can easily extend this to non-Gaussian random fields, such as the exponential distribution, 
\begin{equation}
    p_{\rm exp}(\bm{r} | \bm{\mu}, \bm{Q}) = \sqrt{\frac{\det Q}{2\pi^K}}
        \exp\left(-K\sqrt{\bm{r}^T\bm{Q}\bm{r}}\right)
\end{equation}
and T-distribution,
\begin{equation}
    p_{\rm T}(\bm{r} | \bm{\mu}, \bm{Q}, s) = 
        \frac{\Gamma(\nu/2 + K/2)}{\Gamma(\nu/2)}\sqrt{\frac{\det Q}{(\pi \nu)^K}}
        \left[ 1 + \frac{1}{\nu}\bm{r}^T\bm{Q}\bm{r}\right]^{-(\nu+K)/2},
\end{equation}
where $\nu \geq 1$ denotes the degrees of freedom of the process.

The net effect of the exponential and T-distribution is that the tails of their density function are heavier than those of the GMRF. These heavier tails may be more applicable for sources that have some outliers, such as point-like features, in the image.

\subsection{Higher Order Markov Random Fields}

\begin{figure}[!t]
    \centering
    \includegraphics[width=\linewidth]{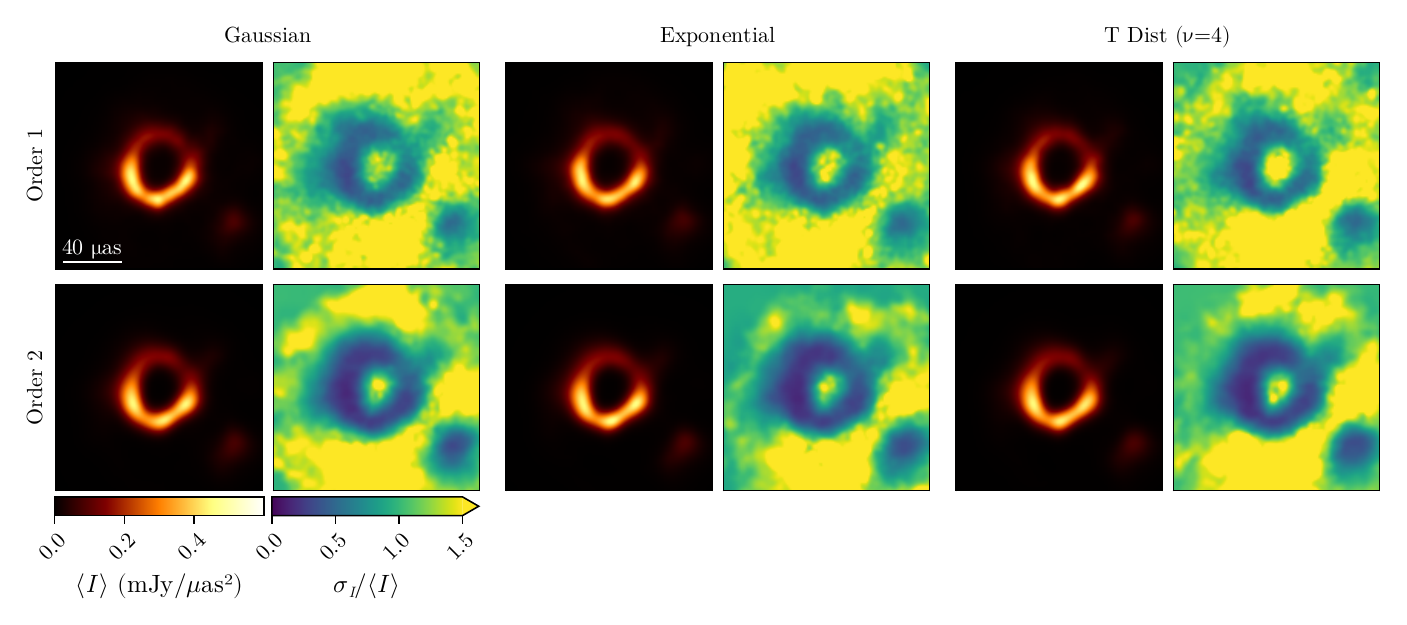}
    \caption{Imaging results on April 6 LO band using different MRF priors. The left group shows the results for a Gaussian base distribution, the middle an exponential distribution, and the right a T distribution with degrees of freedom $\nu = 4$. The top row shows the results for the first-order MRF, and the bottom row shows the second-order MRF.}
    \label{fig:extending_mrf}
\end{figure}

To generalize any Markov Random fields, including the exponential and T distribution MRF, we need to alter the structure of the precision matrix $\bm{Q}$. Following \citet{SPDE2GMRF}, we can easily extend the first-order process by changing the precision matrix to
\begin{equation}
    \bm{Q}_n = \tau(\mathds{1} + \kappa \bm{G})^n,
\end{equation}
where $n$ is a positive integer. In the spectral domain and assuming Gaussian noise and large raster, this process has the power spectrum \citep{SPDE2GMRF}
\begin{equation}
    S(\bm{k})\dd^2\bm{k} \propto \frac{1}{(\kappa^2 + k^2)^n},
\end{equation}
and is a discrete approximation of the integer-order 2D \textit{Matern} Gaussian process, which has autocorrelation function:
\begin{equation}
    C_{\alpha}(r) = \frac{\sigma^2}{2^{\alpha-1}\Gamma(\alpha)}\left(\sqrt{8\alpha}\frac{r}{\rho}\right)^{\alpha}K_\alpha\left(\sqrt{8\alpha}\frac{r}{\rho}\right).
\end{equation}
Here $\alpha = n-1$ and $\sigma$ and $\rho$ are related to $\tau$ and $\kappa$ by
\begin{equation}
\begin{aligned}
    \kappa &= \sqrt{8\alpha}/\rho \\
    \sigma   &= \sqrt{\frac{1}{\tau\kappa^\alpha}\frac{\Gamma(\alpha)}{\Gamma(\alpha+1)4\pi}}.
\end{aligned}
\end{equation}
Increasing the order of the GMRF increases the power-law slope and, thus, the smoothness of the image reconstruction. Note that the first-order GMRF is not an example of the Matern process but remains a valid random field on a grid.

The higher-order random field can be easily included in the exponential and T-distribution random fields by setting $Q = Q_n$. The impact of changing the base distribution and order of the Markov random field is shown in \autoref{fig:extending_mrf} for the 2017 M87 observations on April 6, LO band. In general, \autoref{fig:extending_mrf} demonstrates that the mean reconstructions are nearly identical across the different priors. The fractional uncertainty maps are also very similar across the different priors and orders. However, the second-order MRF reconstructions tend to be smoother, which is expected due to its steeper power spectrum compared to the first-order process.

\newpage

\bibliography{sample631}
\bibliographystyle{aasjournal}



\end{document}